%% file: main.tex
\documentclass[prb,reprint,twocolumn,amsmath,amssymb]{revtex4-1}

\newcommand{\del}{\nabla}

\newcommand{\bvPsi}{\boldsymbol{\varPsi}}

\newcommand{\balpha}{\boldsymbol{\alpha}}
\newcommand{\bb}{\boldsymbol{b}}

\newcommand{\br}{\boldsymbol{\textbf{r}}}

\newcommand{\bv}{\boldsymbol{v}}
\newcommand{\bx}{\boldsymbol{\textbf{x}}}

\newcommand{\bB}{\boldsymbol{\textbf{B}}}

\newcommand{\bH}{\boldsymbol{\textbf{H}}}
\newcommand{\bI}{\boldsymbol{I}}
\newcommand{\bK}{\boldsymbol{\textbf{K}}}

\newcommand{\bM}{\boldsymbol{\textbf{M}}}
\newcommand{\bL}{\boldsymbol{\textbf{L}}}
\newcommand{\bN}{\boldsymbol{N}}

\newcommand{\bR}{\boldsymbol{\textbf{R}}}
\newcommand{\bS}{\boldsymbol{\textbf{S}}}

\newcommand{\bW}{\boldsymbol{\textbf{W}}}

\newcommand{\bPhi}{\boldsymbol{\Phi}}

\newcommand{\bGam}{\boldsymbol{\Gamma}}

\newcommand{\dx}{\,d\bx}

\newcommand{\btH}{\boldsymbol{\tilde{\textbf{H}}}}

\newcommand{\btX}{\boldsymbol{\textbf{X}}}
\newcommand{\btY}{\boldsymbol{\textbf{Y}}}

\newcommand{\op}[1]{\mathcal{#1}}
\newcommand{\order}{\mathcal{O}}

\newcommand{\ket}[1]{\left| #1 \right>} 
\newcommand{\bra}[1]{\left< #1 \right|} 
\newcommand{\braket}[2]{\left< #1 \vphantom{#2} \right | \left. #2 \vphantom{#1} \right>} 
\newcommand{\matrixel}[3]{\left< #1 \vphantom{#2#3} \right| #2 \left| #3 \vphantom{#1#2} \right>} 
\usepackage{graphicx}
\usepackage{graphics}
\usepackage{multirow}
\usepackage{fancyhdr, ifpdf}
\usepackage{amsxtra}
\usepackage{stmaryrd}
\usepackage{mathrsfs}
\usepackage{psfrag}
\usepackage{subfigure}
\usepackage{epsfig}
\usepackage{rotating}
\usepackage{setspace}
\usepackage{float}
\usepackage{amsfonts}
\usepackage{amsbsy}
\usepackage{amscd}
\usepackage{amsthm}
\usepackage{color}
\usepackage{tabularx}
\usepackage{caption}
\usepackage{sidecap}
\usepackage{dcolumn}
\usepackage[ruled]{algorithm2e}
\usepackage{algorithmic}

\definecolor{hellgruen}{rgb}{0.2,0.7,0.2}

\begin{document}
\preprint{}
 \title{A subquadratic-scaling subspace projection method for large-scale Kohn-Sham density functional theory calculations using spectral finite-element discretization}
\author{Phani Motamarri}
\author{Vikram Gavini}
\affiliation{Department of Mechanical Engineering, University of Michigan, Ann Arbor, MI 48109, USA}
\begin{abstract}
\input{abstract.tex}
\end{abstract}
\maketitle
\section{Introduction}\label{sec:intro}
\input{intro}

\section{Kohn-Sham density functional theory}\label{sec:formulation}
\input{formulation.tex}

\section{Mathematical Formulation}\label{sec:math}
\input{math.tex}

\vspace{-0.25in}
\section{Subspace-projection algorithm using finite-element basis}\label{sec:fem}
\input{fem.tex}

\section{Results and discussion}\label{sec:results}
\input{results.tex}
\newpage
\vspace{-0.15in}
\section{Summary}\label{sec:conclusions}
\vspace{-0.1in}
\input{conclusions.tex}
\acknowledgements
\input{acknowledgements.tex}

\appendix \label{sec:appendix}
\input{appendix.tex}

\end{document}

%% file: abstract.tex
We present a subspace projection technique to conduct large-scale Kohn-Sham density functional theory calculations using higher-order spectral finite-element discretization. The proposed method treats both metallic and insulating materials in a single framework, and is applicable to both pseudopotential as well as all-electron calculations. The key ideas involved in the development of this method include: (i) employing a higher-order spectral finite-element basis that is amenable to mesh adaption; (ii) using a Chebyshev filter to construct a subspace which is an approximation to the occupied eigenspace in a given self-consistent field iteration; (iii) using a localization procedure to construct a non-orthogonal localized basis spanning the Chebyshev filtered subspace; (iv) using a Fermi-operator expansion in terms of the subspace-projected Hamiltonian represented in the non-orthogonal localized basis to compute relevant quantities like the density matrix, electron density and band energy. We demonstrate the accuracy and efficiency of the proposed approach on benchmark systems involving pseudopotential calculations on aluminum nano-clusters up to 3430 atoms and on alkane chains up to 7052 atoms, as well as all-electron calculations on silicon nano-clusters up to 3920 electrons. The benchmark studies revealed that accuracies commensurate with chemical accuracy can be obtained with the proposed method, and a subquadratic-scaling with system size was observed for the range of materials systems studied. In particular, for the alkane chains---representing an insulating material---close to linear-scaling is observed, whereas, for aluminum nano-clusters---representing a metallic material---the scaling is observed to be $\order (N^{1.46})$. For all-electron calculations on silicon nano-clusters, the scaling with the number of electrons is computed to be $\order (N^{1.75})$. In all the benchmark systems, significant computational savings have been realized with the proposed approach, with $\sim 10-$fold speedups observed for the largest systems with respect to reference calculations.

%% file: intro.tex
Past few decades have seen an increasingly important role played by electronic structure calculations based on density functional theory in the investigation of materials properties.  The Kohn-Sham approach to density functional theory (DFT) ~\cite{kohn64,kohn65}, in particular, provides a computationally tractable approach to conduct quantum-mechanically informed calculations on ground-state materials properties. The Kohn-Sham approach reduces the many-body problem of interacting electrons into an equivalent problem of non-interacting electrons in an effective mean field that is governed by the electron density. While this formulation has no approximations and is exact in principle for ground-state properties, the quantum mechanical interactions between electrons manifest in the form of an unknown exchange-correlation term in DFT. Various models have been proposed for the exchange-correlation interactions~\cite{XCReview2005}, and these have been shown to predict a wide range of materials properties across various materials systems. Though the Kohn-Sham approach greatly reduces the computational complexity of the many-body Schr$\ddot{\text{o}}$dinger problem, large-scale electronic structure calculations with DFT are still computationally very demanding, and accurate numerical methods that improve the computational efficiency of DFT calculations are desirable.

The widely used numerical implementations of DFT employ a plane-waves basis~\cite{VASP,CASTEP,ABINIT}, which provides an efficient computation of the electrostatic interactions arising in DFT naturally through Fourier transforms. Further, the plane-wave basis provides variational convergence in the ground-state energy with exponential convergence rates. However, the plane-wave basis also suffers from some notable disadvantages. In particular, the simulations are restricted to periodic boundary conditions that are not well suited for material systems containing extended defects (for e.g. dislocations), as well as isolated materials systems such as molecules and nano-clusters. Further, the plane-wave basis does not offer adaptive spatial resolution which can be inefficient in the treatment of certain types of calculations|for e.~g. all-electron calculations, isolated systems|where higher basis resolution is required in some spatial regions and a coarser resolution suffices elsewhere. Moreover, the plane-wave basis functions are extended in real space, which significantly affects the scalability of computations on parallel computing platforms. On the other hand, atomic-orbital-type basis functions~\cite{Pople,Jensen,Qchem,Molpro,SIESTA,FHI-aims}, which constitute other widely employed basis sets in electronic structure calculations, are well suited for isolated systems, but cannot easily handle other boundary conditions. Further, these basis sets do not offer systematic convergence for all materials systems, and due to the non-locality of the basis functions the parallel scalability is significantly affected. Thus, there has been an increasing focus on systematically improvable and scalable real-space techniques for electronic structure calculations over the past decade~\cite{Beck,PARSEC,OCTOPUS,BigDFT,CONQUEST,ONETEP,ACRES,Varga2004,Amartya}.

Among real-space techniques, the finite-element basis~\cite{Brenner-Scott}|a piecewise polynomial basis|presents some key advantages for electronic structure calculations. In particular, the finite-element basis naturally allows for arbitrary boundary conditions, which enables the consideration of isolated, semi-periodic, as well as, periodic materials systems under a single framework. Further, the locality of the basis provides good scalability on parallel computing platforms. Further, the finite-element basis is amenable to adaptive spatial resolution which can effectively be exploited for efficient solution of all-electron DFT calculations~\cite{bylaska,lehtovaara,motam2013} as well as the development of coarse-graining techniques that seamlessly bridge electronic structure calculations with continuum~\cite{QCOFDFT,Bala2010}. There have been significant efforts in the recent past to develop real-space electronic structure calculations based on a finite-element discretization ~\cite{Hermansson,white,tsuchida1995,tsuchida1996,tsuchida1998,pask1999,Batcho2000,pask2001,ACRES,pask2005,Zhou2008,bylaska,lehtovaara,surya2010,Lin2012,Zhou2012,Bao2012,motam2012,Masud2012,motam2013}.

Although the finite-element basis offers some unique advantages for electronic structure calculations, initial studies~\cite{bylaska,Hermansson,surya2010}, which employed linear finite-element basis functions, suggested that they require a large number of linear basis functions|of the order of 100,000 basis functions per atom|to achieve chemical accuracy in electronic structure calculations. This compares poorly with plane-wave basis or atomic-orbital-type basis. A recent investigation~\cite{motam2013} has indicated that the use of adaptive higher-order spectral finite-elements can significantly improve the computational efficiency of real-space electronic structure calculations. In particular, staggering computational savings---of the order of $1000-$fold---relative to linear finite-elements for both all-electron and local pseudopotential calculations have been obtained by using higher-order finite-element discretizations. Further, for accuracies commensurate with chemical accuracy, it was demonstrated that the computational efficiency afforded by higher-order finite-element discretizations is competing with plane-wave discretization for non-periodic  calculations, and is comparable to the Gaussian basis for all-electron calculations to within an order of magnitude. Moreover, the parallel scalability of the finite-element basis was demonstrated on up to 200 processors, where over 90\% parallel efficiency was observed. Using modest computational resources, local pseudopotential calculations had been demonstrated on up to 1688 atoms and all-electron calculations had been demonstrated on up to 600 electrons.

However, we note that the traditional self-consistent approach to solving the discretized nonlinear Kohn-Sham eigenvalue problem involves the diagonalization of the Hamiltonian to obtain orthonormal eigenvectors, the computational complexity of which typically scales as $\order (M\,N^2)$ where $M$ denotes the number of basis functions and $N$ denotes the system size (number of atoms or number of electrons). This cost becomes prohibitively expensive, approaching cubic-scaling, as the system size becomes larger. To this end, numerous efforts have focused on either reducing the prefactor \cite{saad2006,saad2012} associated with the computational cost of the Kohn-Sham DFT calculations  or reducing the computational complexity to have improved-scaling behavior for DFT calculations. The latter methods usually exploit locality in the wavefunctions~\cite{kohn96} directly or indirectly, and can be broadly categorized~\cite{godecker99} into two types: one which calculate the single-electron density matrix, and another which work with its representation in terms of localized Wannier functions. The divide and conquer method~\cite{yang92,Ozaki2006,barrault2007}, Fermi-operator expansion~\cite{godecker93,godecker94,godecker95,baer97}, density-matrix minimization~\cite{vanderbilt93,haynes2006} approach belong to the former category, whereas, the Fermi-operator projection method~\cite{sankey94,stephan98} and the orbital minimization approach~\cite{mauri93,kim95,Gao2009,carlos} belong to the latter category. A comprehensive review of these methods has been provided by G$\ddot{\text{o}}$edecker~\cite{godecker99}, and more recently by Bowler and Miyazaki~\cite{bow2012}. These methods, which rely on the locality of the Wannier functions or the exponential decay of the density matrix in real-space, have been demonstrated to work well for insulating systems, exhibiting linear-scaling with system size. However, for metallic systems, due to the slower decay of the density matrix, the computational complexity of these approaches can deviate significantly, in practice, from linear-scaling. Further, we note that, some of the developed techniques~\cite{vanderbilt93,sankey94,stephan98,carlos} assume the existence of a band-gap, thus restricting these techniques solely to insulating systems. The Fermi-operator expansion method~\cite{godecker93,baer97,godecker99}, which is equally applicable to both insulating and metallic systems, computes the finite-temperature density-matrix through a Chebyshev polynomial approximation of the Fermi distribution function (also referred to as Fermi function) of the Kohn-Sham Hamiltonian. The accuracy of such an expansion depends on the smearing parameter ($\sigma = k_B T$) in the Fermi distribution and the width of the eigenspectrum ($\Delta E$) of the discretized Hamiltonian. In fact, the number of polynomial terms required to achieve a prescribed accuracy~\cite{baer97} is $\order(\frac{\Delta E}{\sigma})$. Numerous recent efforts~\cite{ozaki2007,ceriotti2008,linlin2009,ozaki2010,linlin2013a} have focused towards developing alternate approximations to the Fermi function, or approximations to its spectral representation~\cite{phanish2013a}. A majority of these methods aim to reduce the number of terms used in the expansion to approximate the Fermi function. However, a major drawback with these methods is that they are not efficient for local real-space basis functions like finite-elements, where, typically, more refined discretizations are needed. In a recent study~\cite{motam2013}, it was observed that the width of the eigenspectrum of the finite-element discretized Hamiltonian (using higher-order finite-elements) is $\order(10^3)~Ha$ for pseudopotential calculations, and $\order(10^5)~Ha$ for all-electron calculations. In this article, building on our previous work~\cite{motam2013}, we propose a reduced-scaling subspace projection technique in the framework of spectral finite-element discretization. To this end, we borrow localization ideas from Garcia et al.~\cite{carlos} and develop a subspace iteration technique that treats both metallic and insulating systems on a similar footing. Further, besides pseudopotential calculations, the proposed technique is also applicable to all-electron calculations, as demonstrated in our benchmark studies. The main ideas used in our approach are: (i) employ Chebyshev filtered subspace iteration to compute the occupied eigenspace; (ii) employ a localization procedure to generate non-orthogonal localized wavefunctions spanning the Chebyshev filtered subspace; (iii) use adaptive tolerances to truncate the wavefunctions, with looser tolerances being employed in initial self-consisted field (SCF) iterations and progressively tightening as the SCF iteration approaches convergence, and (iv) employ Fermi-operator expansion in terms of the projected Hamiltonian expressed in the non-orthogonal localized basis to compute the density matrix, electron density and band energy.

We first present an abstract mathematical framework in which the projection of the Hamiltonian into a subspace, corresponding to the occupied eiegenspace, and the associated density matrix are expressed in a non-orthogonal basis spanning the subspace. We then derive expressions for the computation of electron density, constraint on the number of electrons and the band energy in terms of the projected Hamiltonian, which are subsequently used to formulate the subspace projection technique within the framework of finite-element discretization. To this end, the Kohn-Sham Hamiltonian and the corresponding wave-functions are represented in the L{\"o}wdin orthonormalized finite-element basis constructed using spectral finite-elements in conjunction with Gauss-Lobatto-Legendre quadrature rules. The SCF iteration begins with an initial subspace spanned by the localized single-atom wavefunctions, and a Chebyshev filter is applied on this subspace to compute an approximation to the occupied eigenspace. Next, we employ a localization procedure to construct non-orthogonal localized wavefunctions spanning the Chebyshev filtered subspace. The localized wavefunctions are then truncated using a truncation tolerance, below which the localized wavefunctions are set to zero. We note that the proposed approach of providing a compact support for the wavefunctions by using a truncation tolerance on the wavefunctions differs from commonly employed approach of using truncation radius, and presents a more efficient approach as the localized wavefunctions may not necessarily be spherically symmetric about the localization center. The Kohn-Sham Hamiltonian is then projected into the localized basis, and a Fermi-operator expansion in terms of the projected Hamiltonian is employed to compute the finite-temperature density matrix and the electron density. If the truncated wavefunctions are sufficiently sparse, the computational cost of each of the above steps is shown to scale linearly with number of atoms.

The proposed approach is implemented in a parallel computing framework, and the performance of the algorithm is investigated on representative benchmark atomic systems involving metallic aluminum nano-clusters (pseudopotential calculations), insulating alkane chains (pseudopotential calculations), and semi-conducting silicon nano-clusters (all-electron calculations). The scaling behavior is assessed on these materials systems with varying system sizes up to 3430 atoms in the case of aluminum nano-clusters, up to 7052 atoms in the case of alkane chains, and up to 3920 electrons in the case of all-electron silicon nano-clusters. The scaling of the computational time per SCF iteration with the system size is computed to be $\order (N^{1.46})$ for aluminum nano-clusters, $\order (N^{1.18})$ for the alkane chains, and $\order (N^{1.75})$ for the all-electron silicon nano-clusters. One factor contributing to the deviation from linearity is the use of adaptive tolerances---using looser tolerances in the initial SCF iterations and progressively tightening the tolerances as the SCF approaches convergence---which partially sacrifices the scaling for good accuracy in the ground-state energies. We note that the computed ground-state energies using the proposed approach are within 5 $meV$ per atom for pseudopotential calculations and are within 5 $mHa$ per atom for all-electron calculations with respect to the reference ground-state energies. Further, in our implementation, we switch from using sparse data-structures to dense data-structures when the density fraction of the localized wavefunctions exceeds $2\%$, as the computational cost of using parallel sparse data-structures is observed to exceed that of dense data-structures beyond this point. This is another factor contributing to the deviation from linear-scaling as discussed in the Appendix. Our results suggest that significant computational savings can be realized using the proposed approach, where $\sim 10-$fold speedups are obtained with respect to reference benchmark calculations for the largest systems.

The remainder of the paper is organized as follows. Section II describes the real-space formulation of the Kohn-Sham DFT problem, followed by the presentation of the mathematical formalism in Section III.  Section IV describes the various steps involved in the subspace projection technique within the framework of spectral finite-element discretization. Section V presents the numerical study on three representative materials systems demonstrating the accuracy, computational efficiency and scaling of our approach. We finally conclude with a summary and outlook in Section VI.

%% file: formulation.tex
The variational problem  of evaluating the ground-state properties in density functional theory for a materials system consisting of $N_a$ nuclei is equivalent to solving the following non-linear Kohn-Sham eigenvalue problem~\cite{kohn65}:
\begin{equation}\label{contEigen}
\left(-\frac{1}{2} \del^{2} + V_{\text{eff}}(\rho,\bR)\right) {\psi}_{i} = \epsilon_{i} {\psi}_{i},\;\;\;\;\; i = 1,2,\cdots
\end{equation}
where
$\epsilon_{i}$ and $\psi_{i}$ denote the eigenvalues and corresponding eigenfunctions (also referred to as the canonical wavefunctions) of the Hamiltonian, respectively. We denote by $\bR = \{\bR_1,\,\bR_2,\,\cdots \bR_{N_a}\}$ the collection of all nuclear positions in the materials system. In the present work, we restrict ourselves to a non-periodic setting and present the formulation for this case. However, we note that the ideas presented in the work can easily be generalized to periodic or semi-periodic materials systems. Further, for the sake of simplicity, we present the formulation for a spin independent Hamiltonian, and note that the extension to spin-dependent Hamiltonians~\cite{rmartin} follows along similar lines.
The electron density in terms of the canonical wavefunctions is given by
\begin{equation}
\rho(\bx) = 2\sum_{i}f(\epsilon_i,\mu)|\psi_i(\bx)|^2 \,\,,
\end{equation}
where $f(\epsilon_i,\mu)$ is the orbital occupancy function, whose range lies in the interval $\left[0,1\right]$, and $\mu$ represents the Fermi-energy. We note that the factor 2 in the above equation represents the case of a spin independent system, where each orbital is occupied by two electrons. In ground-state calculations, the orbital occupancy function $f(\epsilon,\mu)$ is given by the Heaviside function
\begin{equation}
f(\epsilon,\mu) = \begin{cases}
1 & \text{if $\epsilon < \mu$},\\
0 & \text{otherwise}\,\,.
\end{cases}
\end{equation}
However, it is common in density functional theory calculations to represent $f$ by the Fermi distribution~\cite{VASP,godecker99} given by
\begin{equation}\label{fermidirac1}
f(\epsilon,\mu) = \frac{1}{1 + \exp\left(\frac{\epsilon - \mu}{\sigma} \right)}\,,
\end{equation}
where $\sigma$ is a smearing parameter. We note that as $\sigma\to 0$, the Fermi distribution tends to the Heaviside function. Such a smearing of the orbital occupancy function avoids numerical instabilities that may arise in the solution of the non-linear Kohn-Sham eigenvalue problem, especially in materials systems where there are a large number of eigenstates around the Fermi energy. Often, the numerical value of $\sigma$ is chosen to be $\sigma=k_{B}T$, where $k_B$ denotes the Boltzmann constant and $T$ denotes the finite temperature used for smearing the orbital occupancy function.
The Fermi-energy $\mu$ is computed from the constraint on the total number of electrons in the system ($N_e$) given by
\begin{equation}\label{cons}
\int \rho(\bx) \dx = 2\sum_i f(\epsilon_i,\mu) = N_e\,.
\end{equation}
Henceforth, we denote by $f$ the finite-temperature Fermi distribution given in equation~\eqref{fermidirac1}.

The effective single-electron potential, $V_{\text{eff}}(\rho,\bR)$, in the Hamiltonian in equation~\eqref{contEigen} is given by
\begin{equation}\label{veff}
 V_{\text{eff}}(\rho,\bR) = V_{\text{xc}}(\rho) +V_{\text{H}}(\rho) +V_{\text{ext}}(\bR)\,.
\end{equation}
In the above, $V_{\text{xc}}(\rho)$ denotes the exchange-correlation potential that accounts for quantum-mechanical interactions between electrons, and is given by the first variational derivative of the exchange-correlation energy $E_{\text{xc}}$\,:
\begin{equation}
V_{\text{xc}}(\rho) = \frac{\delta E_{\text{xc}}}{\delta \rho}\,.
\end{equation}
In this work, we adopt the local-density approximation (LDA)~\cite{rmartin} for the exchange-correlation functional. Other approximations such as the local spin density approximation (LSDA)~\cite{rmartin} and generalized gradient approximation (GGA)~\cite{gga1,gga2} can be incorporated into the present formulation in a straightforward manner. In the case of LDA~\cite{alder,perdew}, the exchange-correlation energy is given by
\begin{equation}\label{exc}
E_{\text{xc}}(\rho) = \int \varepsilon_{\text{xc}}(\rho)\rho(\bx) \dx\,\,,
\end{equation}
where $\varepsilon_{\text{xc}}(\rho)=\varepsilon_x(\rho)+\varepsilon_c(\rho)$, and
\begin{equation}
\varepsilon_x(\rho) = -\frac{3}{4}\left(\frac{3}{\pi}\right)^{1/3}\rho^{1/3}(\bx) \,\,,
\end{equation}
\begin{equation}\label{corr}
\varepsilon_c(\rho) = \begin{cases}
&\frac{\gamma}{(1 + \beta_1\sqrt(r_s) + \beta_2r_s)}\;\;\;\;\;\;\;\;\;\;\;\;\;\;\;\;\;\;\;\;\;\;\;r_s\geq1,\\
&A\,\log r_s + B + C\,r_s\log r_s + D\,r_s\;\;\;\;\;\;\;\;r_s\,<\,1,
\end{cases}
\end{equation}
and $r_s = (3/4\pi\rho)^{1/3}$. Specifically, we use the Ceperley and Alder constants  \cite{perdew} in equation~\eqref{corr}.
The remainder of the effective single-electron potential accounts for the electrostatic interactions. In particular, $V_{\text{H}}(\rho)$ (Hartree potential) denotes the classical electrostatic potential corresponding to the electron density distribution and is given by
\begin{equation}
V_\text{H}(\rho) = \int \frac{\rho(\bx')}{|\bx - \bx'|} \dx'\,.
\end{equation}
$ V_{\text{ext}}(\bx,\bR) $ denotes the external electrostatic potential corresponding to the nuclear charges, and is given by
\begin{equation}\label{extpot}
V_{\text{ext}}(\bx,\bR)  = -\sum_{I=1}^{N_a} \frac{Z_I}{|\bx - \bR_I|} \,,
\end{equation}
with $Z_{I}$ denoting the atomic number of the $I^{th}$ nucleus in the given materials system. The tightly bound core electrons close to the nucleus of an atom may not influence the chemical bonding, and may not play a significant role in governing many material properties. Hence, it is a common practice to adopt the pseudopotential approach, where only the wavefunctions for the valence electrons are computed. The pseudopotential, which provides the effective electrostatic potential of the nucleus and core electrons, is often defined by an operator
\begin{equation}
\op {V_{\text{PS}}} = \op {V_{\text{loc}}} + \op {V_{\text{nl}}}\,,
\end{equation}
where $\op {V_{\text{loc}}}$ is the local part of the pseudopotential operator and $\op {V_{\text{nl}}}$ is the non-local part of the operator. In this work, we use the norm-conserving Troullier-Martins pseudopotential~\cite{tm91} in the Kleinman-Bylander form~\cite{bylander82}. The action of these operators on a Kohn-Sham wavefunction in the real-space is given by
\begin{align}
V_{\text{loc}}(\bx,\bR) \psi(\bx) &= \sum_{J=1}^{N_a} V^{J}_{\text{loc}}(\bx-\bR_J)\psi(\bx)\label{Local_PS}\,,\\
V_{\text{nl}}(\bx,\bR)\psi(\bx) &= \sum_{J = 1}^{N_a} \sum_{lm} C^{J}_{lm}V^{J}_{lm}\phi^{J}_{lm}(\bx-\bR_J)\Delta V^{J}_{l}(\bx-\bR_J)\,,\label{NL_PS}
\end{align}
where
\begin{gather*}
\Delta V^{J}_{l}(\bx - \bR_J) = V^{J}_{l}(\bx - \bR_J) - V^{J}_{\text{loc}}(\bx - \bR_J)\,,\\
C^{J}_{lm} = {\int \phi^{J}_{lm}(\bx-\bR_J) \Delta V^{J}_{l}(\bx-\bR_J) \psi(\bx) \dx}\,,\\
\frac{1}{V^{J}_{lm}} = {\int \phi^{J}_{lm}(\bx-\bR_J) \Delta V^{J}_{l}(\bx-\bR_J) \phi^{J}_{lm}(\bx - \bR_J)\dx}\,\,.
\end{gather*}
In the above, $V^{J}_{l}(\bx - \bR_J)$ denotes the pseudopotential component of the atom $J$ corresponding to the azimuthal quantum number $l$, $V^{J}_{\text{loc}}(\bx-\bR_J)$ is the corresponding local potential, and $\phi^{J}_{lm}(\bx-\bR_J)$ is the corresponding single-atom pseudo-wavefunction with azimuthal quantum number $l $ and magnetic quantum number $m$. We note that the computation of the Hartree potential ($V_\text{H}$), the external potential ($V_{\text{ext}}$) in the case of all-electron calculations, or the local part of pseudopotential ($V_{\text{loc}}$), is extended in real-space. However, noting that the kernel corresponding to the extended interactions is the Green's function of the Laplace operator, these quantities can be efficiently computed by taking recourse to the solution of a Poisson problem. We refer to Suryanarayana et al.~\cite{surya2010} and Motamarri et al.~\cite{motam2013} for details of the local reformulation of the extended electrostatic interactions.

Finally, for given positions of nuclei, the system of equations corresponding to the Kohn-Sham eigenvalue problem are:
\begin{subequations}\label{ksproblem}
\begin{gather}
\left(-\frac{1}{2} \del^{2} +  V_{\text{xc}}(\rho) + V_{\text{H}}(\rho) + V_{\text{ext}}(\bR) \right){\psi}_{i} = \epsilon_{i} {\psi}_{i},\\
2\sum_{i}f(\epsilon_i,\mu) = N_e,\\
\rho(\bx) = 2\sum_{i}f(\epsilon_i,\mu)|\psi_{i}(\bx)|^2.
\end{gather}
\end{subequations}
In the case of all-electron calculations, $V_{\text{ext}}$ denotes the Coulomb-singular potential corresponding to all nuclei, and further, all the wavefunctions, including those of the core electrons, are computed. In the case of pseudopotential calculations, $V_{\text{ext}}=V_{\text{loc}}+V_{\text{nl}}$ and only the wavefunctions corresponding to the valence electrons are computed. Furthermore, the formulation in ~\eqref{ksproblem} represents a nonlinear eigenvalue problem which has to be solved self-consistently. Upon self-consistently solving~\eqref{ksproblem}, the ground-state energy of the system is given by
\begin{align}
E_{\text{tot}} =\,& E_{\text{band}} + \int (\varepsilon_{\text{xc}}(\rho) -V_{\text{xc}}(\rho))\, \rho \dx  \notag\\
&  -\frac{1}{2}\int \rho V_{\text{H}}(\rho) \dx + E_{\text{ZZ}}\,,
\end{align}
where $E_{\text{band}}$, denoting the band energy, is given by
\begin{equation}
E_{\text{band}} = 2 \sum_{i} f(\epsilon_i,\mu) \epsilon_i\,,
\end{equation}
and $E_{\text{ZZ}}$, denoting the nuclear-nuclear repulsive energy, is given by
\begin{equation}
E_{\text{ZZ}} = \sum_{I,J=1 \atop I \neq J \,,\, I < J}^{N_a} \frac{Z_I Z_{J}}{|\bR_{I} - \bR_{J}|}\,.
\end{equation}
In the above, $Z_I$ denotes the valence charge of the $I^{th}$ nucleus in the case of pseudopotential calculations, and denotes the atomic number in the case of all-electron calculation.

%% file: math.tex
In this section, we discuss some mathematical preliminaries and present the expression for the projection of the Hamiltonian operator into a subspace spanned by a non-orthogonal basis. Subsequently, we derive the density matrix corresponding to the projected Hamiltonian, and present the expressions for the computation of electron density, constraint on the number of electrons and the band energy in terms of the projected Hamiltonian. This constitutes the mathematical formulation for the subspace projection technique within the framework of spectral finite-element discretization, described in the next section.

Let $\op H $ denote the Hermitian operator representing the Hamiltonian of interest defined on the infinite dimensional Hilbert space $\mathbb{H}$. We note that $\mathbb{H}$ is a space of functions  equipped with the inner product $\left<.|.\right>$ and a norm $\parallel. \parallel$ derived from the inner product. Let $\mathbb{V}_h^{M}\subset \mathbb{H}$ be the finite-dimensional subspace of $\mathbb{H}$ with dimension $M$, and $\{\ket {q_i}\}$ denote an orthonormal basis for $\mathbb{V}_h^{M}$. In the present case, the representation of $\ket {q_i}$ in the real space, $\braket{\bx}{q_i}=q_i(\bx)$, denotes the L{\"o}wdin orthonormalized finite-element basis employed in this study, as discussed subsequently in section~\ref{sec:fem}.  We define the projection operator into the subspace $\mathbb{V}_h^{M}$ to be $\op P^q:\mathbb{H} \rightarrow \mathbb{V}_h^{M}$ given by
\begin{equation}\label{projq}
\op P^q = \sum_{i=1}^{M}\ket{q_i}\bra{q_i}\,.
\end{equation}
The projection of the Hamiltonian into $\mathbb{V}_h^{M}$ is given by $\op P^q \op H: \mathbb{V}_h^{M}\rightarrow \mathbb{V}_h^{M}$, or equivalently $\op P^q \op H \op P^q: \mathbb{H} \rightarrow \mathbb{V}_h^{M}$. We denote the operator corresponding to the projected Hamiltonian to be $\op {\tilde{H}} \equiv \op P^q \op H \op P^q$, and the matrix corresponding to $\op{\tilde{H}}$  expressed in $\{\ket {q_i}\}$ basis by $\tilde{\bH}$ with the matrix element given by $\tilde{\text{H}}_{ij} = \matrixel {q_i} {\op {\tilde{H}}} {q_j}$.

\subsection{Projection of the Hamiltonian into a non-orthogonal basis}
\vspace{-0.15in}
We consider a subspace $\mathbb{V}^{N} \subset \mathbb{V}_h^{M}$, which approximates the occupied eigenspace of $\op {\tilde{H}}$ that can be computed, for instance, using a Chebyshev filtering approach~\cite{saad2006,motam2013} (discussed subsequently in section~IV). Let $\{ \ket {\phi_{\alpha}}\}$ represent a non-orthonormal basis which spans $\mathbb{V}^{N}$. We denote by $\op P^\phi:\mathbb{V}_h^{M} \rightarrow \mathbb{V}^{N}$ the projection operator into the space $\mathbb{V}^{N}$, and is given by
\begin{equation}\label{projphi}
\op P^\phi = \sum_{\alpha,\beta=1}^{N}\ket{\phi_\alpha}S^{-1}_{\alpha\beta}\bra{\phi_\beta}\,,
\end{equation}
where $\bS$ denotes the overlap matrix with matrix elements $S_{\alpha \beta} = \braket {\phi_\alpha} {\phi_\beta}$. We denote the projection of $\op {\tilde{H}}$ into $\mathbb{V}^{N}$ by $\op H^{\phi}$, and is given by $\op H^{\phi} \equiv\op P^\phi \op {\tilde{H}} \op P^\phi: \mathbb{V}_h^{M} \rightarrow \mathbb{V}^{N}$. Denoting the matrix corresponding to $\op H^{\phi}$ expressed in $\{\ket {q_i}\}$ basis to be $\bH^{q}$ and in $\{\ket {\phi_{\alpha}}\}$ basis to be $\bH^{\phi}$, we derive the expressions for the corresponding matrix elements $\text{H}^{q}_{ij}$ and $\text{H}^{\phi}_{\alpha \beta}$ using equations~\eqref{projq} and ~\eqref{projphi} as follows:
\begin{align} \label{compRij}
\text{H}^{q}_{ij} &=  \matrixel {q_i} {\op H^{\phi}} {q_j}
= \matrixel {q_i} {\op {P^\phi} \op {\tilde{H}} \op {P^\phi}} {q_j}  \nonumber \\
&=\sum_{\alpha,\beta = 1 \atop \gamma,\delta = 1}^{N}  \braket {q_i} {\phi_\alpha} S^{-1}_{\alpha\beta} \matrixel {\phi_\beta} {\op {\tilde{H}}} {\phi_\gamma} S^{-1}_{\gamma\delta} \braket {\phi_\delta} {q_j}\,.
\end{align}
Since $\op P^q$ is idempotent ($\op P^q \op P^q = \op P^q$), we note that $\op {\tilde{H}} = \op P^q \op {\tilde{H}} \op P^q$. Hence, equation~\eqref{compRij} can be written as
\begin{align}
\text{H}^{q}_{ij}&= \sum_{\alpha,\beta = 1 \atop \gamma,\delta = 1}^{N} \braket {q_i} {\phi_\alpha} S^{-1}_{\alpha\beta} \matrixel {\phi_\beta} {\op P^q \op {\tilde{H}} {\op P^q}} {\phi_\gamma} S^{-1}_{\gamma\delta} \braket {\phi_\delta} {q_j} \nonumber \\
&=\sum_{\alpha,\beta = 1 \atop \gamma,\delta = 1}^{N}\;\sum_{ k ,\, l = 1 }^{M} \phi_{i\alpha}\, S^{-1}_{\alpha\beta}\,
 \phi_{k\beta}^{*} \,\tilde{\text{H}}_{kl}\,\phi_{l \gamma}\,S^{-1}_{\gamma \delta}\, \phi_{j\delta}^{*}\, ,\,
\end{align}
where $\phi_{i\alpha} = \braket {q_i} {\phi_\alpha}$ and $\phi_{k \beta}^{*}$ denotes the complex conjugate of $\phi_{k \beta}$.
The above equation can be conveniently recast in terms of matrices as:
\begin{equation}\label{projHamq}
\bH^{q} = \bPhi \bPhi^{+} \btH \bPhi \bPhi^{+}
\end{equation}
where $\bPhi$ denotes a matrix whose column vectors are the components of $\ket{\phi_\alpha}$ in $\ket {q_i}$ basis, and $\bPhi^{+}$ denotes the Moore-Penrose pseudo-inverse of $\bPhi$ given by
\begin{equation}
\bPhi^{+} = \bS^{-1} \bPhi^{\dagger}
\end{equation}
with $\bPhi^{\dagger}$ denoting the conjugate transpose of the matrix $\bPhi$.
We now derive the expression for the matrix element corresponding to the operator $\op H^{\phi}$ expressed in the non-orthonormal basis $\{ \ket {\phi_{\alpha}}\}$:
\begin{align}\label{compRalpbet}
\text{H}^{\phi} _{\alpha \beta} & = \sum_{\gamma = 1} ^{N}  S^{-1}_{\alpha \gamma} \matrixel {\phi_\gamma} {\op H^{\phi}}  {\phi_\beta}  =\sum_{\gamma = 1} ^{N}  S^{-1}_{\alpha \gamma} \matrixel {\phi_\gamma} {\op {P^\phi} \op {\tilde{H}} \op {P^\phi}}  {\phi_\beta} \nonumber \\
&=\sum_{\gamma = 1} ^{N}  S^{-1}_{\alpha \gamma} \matrixel {\phi_\gamma} {\op {\tilde{H}}}  {\phi_\beta}
=\sum_{\gamma = 1} ^{N}  S^{-1}_{\alpha \gamma} \matrixel {\phi_\gamma} {\op P^q \op {\tilde{H}} \op P^q}  {\phi_\beta} \nonumber \\
&=\sum_{\gamma = 1} ^{N}\; \sum_{ k,\, l = 1 }^{M}  S^{-1}_{\alpha \gamma} \braket {\phi_\gamma} {q_k}  \matrixel {q_k} {\op {\tilde{H}}} {q_l} \braket {q_l} {\phi_\beta} \nonumber\\
&=\sum_{\gamma = 1} ^{N}\; \sum_{ k,\,  l = 1 }^{M}  S^{-1}_{\alpha \gamma} \,\phi_{k\gamma}^{*} \,\tilde{\text{H}}_{kl}\,\phi_{l \beta}\,.
\end{align}
Using matrices, the above equation~\eqref{compRalpbet} can be conveniently expressed as
\begin{equation}\label{projHamalp}
\bH^{\phi} =  \bPhi^{+} \btH \bPhi\,.
\end{equation}
We also note the following relation between $\bH^{\phi}$ and $\bH^{q}$ using equations~\eqref{projHamq} and ~\eqref{projHamalp}:
\begin{equation}
\bH^{q} =\bPhi \bH^{\phi} \bPhi^{+}\,.
\end{equation}
\vspace{-1.2in}
\subsection{Density Matrix}
We now consider the single particle density operator ($\varGamma$) corresponding to $\op H^{\phi}$ given by
\begin{align}\label{denop}
\varGamma &= \sum_{i=1}^{N} f(\epsilon^{\phi}_i) \ket {\psi_i^{\phi}} \bra {\psi_i^{\phi}} \nonumber \\
&=f(\op H^{\phi})\,,
\end{align}
where $\epsilon^{\phi}_i$ and $\ket {\psi_i^{\phi}}$ denote the eigenvalues and the corresponding eigenvectors of $\op H^{\phi}$, and we note that the relation $\varGamma = f(\op H^{\phi}) $ follows from the spectral decomposition of the Hermitian operator $\op H^{\phi}$. Denoting the matrix representation of $\varGamma$ in $\{ \ket {q_i} \}$ basis by $\bGam$, we have the following relation between the matrices  $\bGam$ and $\bH^{q}$ from equation~\eqref{denop}:
\begin{equation}\label{denmat}
\bGam = f(\bH^{q})\,.
\end{equation}
We now derive the expression for $\bGam$ in terms of the matrices $\bH^{\phi}$ and $\bPhi$. To this end, we note that the function $f(\epsilon)$ represents the Fermi distribution (equation~\eqref{fermidirac1}), which is an analytic function. Hence, this admits a power series representation given by
\begin{equation}
f(\bH^{q}) = \sum_{k=0}^{\infty} a_k \left({\bH^{q}}\right)^{k} \,.
\end{equation}
Using equation~\eqref{projHamq} and the relation $\bPhi^{+} \bPhi = \bI_N$, where $\bI_N$ denotes an identity matrix of dimension  $N$, we have
\begin{align} \label{denmat}
\bGam&= \sum_{k=0}^{\infty} a_k \left({\bH^{q}}\right)^{k} =  \sum_{k=0}^{\infty} a_k\;(\bPhi \bPhi^{+} \btH \bPhi \bPhi^{+})^{k} \nonumber \\
&= \sum_{k=0}^{\infty} a_k\;\bPhi (\bPhi^{+} \btH \bPhi )^{k} \bPhi^{+}= \bPhi\left(\sum_{k=0}^{\infty}   a_k (\bH^{\phi})^{k}\right)
 \bPhi^{+}\nonumber\\
&= \bPhi f(\bH^{\phi}) \bPhi^{+}\,.
\end{align}
We note that the electron density $\rho(\bx)$ is related to the diagonal of the density operator $\varGamma$ expressed in the real space, and is given by
\begin{equation}\label{elecden}
\rho(\bx) = 2 \,\matrixel {\bx} {\varGamma}{\bx} = 2 \sum_{i,j=1}^{M} \Gamma_{ij}\,q_i(\bx)\,q_j(\bx)\,,
\end{equation}
where $\Gamma_{ij}$ denote the matrix elements of $\bGam$. We note that factor 2 in the above equation represents the case of a spin independent system, where each orbital is occupied by two electrons. We now derive an expression for the constraint on the total number of electrons as given in equation~\eqref{cons}. The total number of electrons in the materials system is given by
\begin{align}
\int \rho(\bx) \dx &=  2\;\int \sum_{i, j = 1 }^{M} \Gamma_{ij}\,q_i(\bx)\,q_j(\bx) \dx  = 2\;\sum_{i=1}^{M} \Gamma_{ii}\nonumber \\
&= 2\; \text{tr}\left(\bGam\right) = 2\;\text{tr}\left( \bPhi f(\bH^{\phi}) \bPhi^{+}\right) \nonumber \\[0.05in]
&=2\;\text{tr}\left(f(\bH^{\phi}) \bPhi^{+}\bPhi\right) = 2\;\text{tr}\left(f(\bH^{\phi})\right)\,.
\end{align}
Hence, the constraint on the total number of electrons in  equation~\eqref{cons} can be reformulated as
\begin{equation}\label{cons1}
2\;\text{tr}\left(f(\bH^{\phi})\right) = N_e\,.
\end{equation}
Finally, the band-energy ($E_b$), which is required in the calculation of the ground-state energy, is given by
\begin{align}\label{bandenergy}
E_b &= 2\;\text{tr} \left(\bGam\, \bH^{q}\right)=2\;\text{tr} \left( \bPhi f(\bH^{\phi}) \bPhi^{+}  \bPhi \bH^{\phi} \bPhi^{+}\right) \nonumber \\
&=2\;\text{tr} \left( \bPhi f(\bH^{\phi}) \bH^{\phi} \bPhi^{+}\right) =  2\;\text{tr} \left(  \bPhi^{+}\bPhi f(\bH^{\phi}) \bH^{\phi}\right)\nonumber \\
&= 2\; \text{tr} \left( f(\bH^{\phi}) \bH^{\phi}\right)\,.
\end{align}

%% file: fem.tex
In this section, we introduce the finite-element discretization of the Kohn Sham eigenvalue problem along the lines of our prior work~\cite{motam2013}, and subsequently present the subspace projection algorithm used to reduce the computational complexity involved in the solution of the Kohn-Sham problem.
\vspace{-0.15in}
\subsection{Discrete Kohn-Sham eigenvalue problem}
\vspace{-0.15in}
As denoted in the previous section, let $\mathbb{V}_h^{M}$ represent the finite-dimensional subspace with dimension $M$. In particular, we consider the finite-dimensional space spanned by the finite-element basis, which is a piecewise polynomial basis generated from a finite-element discretization~\cite{Brenner-Scott} whose characteristic mesh-size is denoted by $h$. The representation of the various fields in the Kohn-Sham eigenvalue problem~\eqref{ksproblem}---the wavefunctions and the electrostatic potential---in the finite-element basis is given by
\begin{equation}\label{fem}
\psi^{h}_{i}(\bx) = \sum_{j=1}^{M} N^{h}_j(\bx) \psi^{j}_{i}\,\,,
\end{equation}
\begin{equation}\label{fem_phi}
\phi^{h}(\bx) = \sum_{j=1}^{M} N^{h}_j(\bx) \phi^{j}\,\,,
\end{equation}
where $N^{h}_{j}:1\leq j \leq M$ denotes the finite-element basis spanning $\mathbb{V}_h^{M}$. We note that $\psi^{h}_{i}$ and $\phi^{h}$ denote the finite-element discretized fields, with $\psi^{j}_{i}$ and $\phi^{j}$ denoting the coefficients in the expansion of the $i^{th}$ discretized wavefunction and electrostatic potential, respectively. Further, from the Kronecker delta property~\cite{Brenner-Scott} of the finite-element basis, $\psi^{j}_{i}$ and $\phi^{j}$ also correspond to the nodal values of the respective fields at the $j^{th}$ node on the finite-element mesh. We note that the finite-element basis is a non-orthogonal basis, and, thus, the discretization of the Kohn-Sham eigenvalue problem~\eqref{ksproblem} results in a generalized eigenvalue problem given by
\begin{equation}\label{ghep}
\bH \hat{\bvPsi}_{i} = \epsilon^{h}_{i} \bM \hat{\bvPsi}_{i}\,\,,
\end{equation}
where $\bH$  denotes the discrete Hamiltonian matrix with matrix elements $\text{H}_{jk}$, $\bM$ denotes the overlap matrix (or commonly referred to as the mass matrix in finite-element literature) with matrix elements $\text{M}_{jk}$, and $\epsilon^{h}_{i}$ denotes the $i^{th}$ eigenvalue corresponding to the discrete eigenvector $\hat{\bvPsi}_i$. The expression for the discrete Hamiltonian matrix $\text{H}_{jk}$ in a non-periodic setting with $\mathbb{H}=H^1_0(\Omega)$ (space of functions with compact support on $\Omega$) is given by
\begin{equation}\label{discreteHam}
H_{jk} = H_{jk}^{\text{loc}} + H_{jk}^{\text{nl}}\,,
\end{equation}
where
\begin{eqnarray}
\text{H}_{jk}^{\text{loc}} = &&\frac{1}{2} \int \del N_{j} (\bx) .\del N_k(\bx) \dx \nonumber\\ &+&\int V^{h}_{\text{eff,loc}}(\bx,\bR) N_{j}(\bx) N_{k}(\bx) \dx \,\,.
\end{eqnarray}
In the above, $V_{\text{eff,loc}}^{h}$ denotes the local part of the effective single-electron potential computed in the finite-element basis (the discretized effective single-electron potential). In the case of pseudopotential calculations, $V^h_{\text{eff,loc}}=V^h_{\text{xc}}+V^h_{\text{H}}+V^h_{\text{loc}}$, where $V^h_{\text{loc}}$ denotes the discretized local part of the pseudopotential (cf.~equation~\eqref{Local_PS}). In the case of all-electron calculations, $V^h_{\text{eff,loc}}=V^h_{\text{xc}}+V^h_{\text{H}}+V^h_{\text{ext}}$ with $V^h_{\text{ext}}$ denoting the discretized all-electron external Coulomb potential from all nuclei. As noted in section~\ref{sec:formulation}, $V^h_{\text{H}}$ and $V^h_{\text{ext}}$ are computed as solutions of the Poisson's equation in the finite-element basis. In the case of all-electron calculations, the term $H_{jk}^{nl}$ in ~\eqref{discreteHam} is zero, while in the case of pseudopotential calculations it is given by
\begin{equation}
H_{jk}^{nl} = \sum_{J = 1}^{N_a} \sum_{lm} C^{J}_{lm,j}V^{J}_{lm}C^{J}_{lm,k}\,,
\end{equation}
where
\begin{equation}
C^{J}_{lm,j} =  \int \phi^{J}_{lm}(\bx-\bR_J) \Delta V^{J}_{l}(\bx-\bR_J) N_j(\bx) \dx\,.
\end{equation}
Finally, the matrix elements of the overlap matrix $\bM$ are given by $\text{M}_{jk}=\int N_j(\bx) N_k(\bx) \dx$.
We note that the matrices $\bH^{\text{loc}}$ and $\bM$ are sparse as the finite-element basis functions are local in real space and have a compact support (a finite region where the function is non-zero and zero elsewhere). Further, the vectors $C^{J}_{lm,j}$ in $\bH^{nl}$ are also sparse since both $\phi^{J}_{lm}(\bx-\bR_J)$ and $ \Delta V^{J}_{l}(\bx-\bR_J)$ have a compact support, thus rendering a sparse structure to the discrete Hamiltonian $\bH$.

It is now desirable to transform the generalized eigenvalue problem~\eqref{ghep} into a standard eigenvalue problem for which a wide range of efficient solution procedures are available. Since the matrix $\bM$ is positive definite symmetric, there exists a unique positive definite symmetric square root of $\bM$, and is denoted by $\bM^{1/2}$. Hence, the following holds true:
\begin{align}
\bH \hat{\bvPsi}_{i} &= \epsilon^{h}_{i} \bM \hat{\bvPsi}_{i} \nonumber \\
\Rightarrow\qquad \bH \hat{\bvPsi}_{i} &= \epsilon^{h}_{i} \bM^{1/2} \bM^{1/2} \hat{\bvPsi}_{i} \nonumber \\
\Rightarrow\qquad \btH \tilde{\bvPsi}_{i} &= \epsilon^{h}_{i} \tilde{\bvPsi}_{i} \label{hep}\,,
\end{align}
where
\begin{align*}
\tilde{\bvPsi}_{i} &= \bM^{1/2} \hat{\bvPsi}_{i}\\
\btH &= \bM^{-1/2}\bH\bM^{-1/2}\,.
\end{align*}
We note that  $\btH$ is a Hermitian matrix, and~\eqref{hep} represents a standard Hermitian eigenvalue problem. The actual eigenvectors are recovered by the transformation $\hat{\bvPsi}_{i} = \bM^{-1/2} \tilde{\bvPsi}_{i}$. We remark that $\tilde{\bvPsi}_{i}$ is a vector containing the expansion coefficients of the discretized eigenfunction $\psi^{h}_i(\bx)$ expressed in L{\"o}wdin orthonormalized finite-element basis $q_j(\bx): {1\leq j \leq M}$ spanning the finite-element space. We note the following relation between $q_j(\bx)$ and $N_j(\bx)$:
\begin{equation}\label{orthfem}
q_j(\bx) = \sum_{k=1}^{M} M_{jk}^{-1/2} N_k(\bx)\,.
\end{equation}
Furthermore, we note that the transformation to a standard eigenvalue problem~\eqref{hep} is computationally advantageous if the matrix $\bM^{-1/2}$ can be evaluated with modest computational cost and the matrix $\btH$ has the same sparsity structure as the matrix $\bH$. This is immediately possible by using a spectral finite-element basis in conjunction with the use of Gauss-Lobatto-Legendre (GLL) quadrature for the evaluation of integrals in the overlap matrix, that renders the overlap matrix diagonal~\cite{motam2013}. This has been employed in Motamarri et al.~\cite{motam2013} to develop an efficient and scalable approach using adaptive higher-order spectral finite-element discretization of the Kohn-Sham DFT problem. Materials systems as large as 1688 atom aluminum nano-clusters have been simulated, and the parallel scalability of the algorithms has been demonstrated up to 200 processors~\cite{motam2013}. In Motamarri et al.~\cite{motam2013}, the self-consistent field (SCF) iteration consists of employing a Chebyshev filtering approach~\cite{saad2006} to compute the occupied eigenspace of the spectral finite-element discretized Hamiltonian $\btH$. The computational complexity of the Chebyshev filtering scales as $\order(M\,N)$, where $N$ denotes the number of eigenstates of interest. Upon computing the approximate eigenspace, the Chebyshev filtered vectors spanning the eigenspace are orthonormalized, and the projection of the Hamiltonian into this orthonormal basis is computed. The computational complexity of this orthonormalization and projection scales as $\order(M\,N^2)$. Finally, the projected Hamiltonian is diagonalized to compute the eigenvalues and eigenvectors, which, in turn, are used in the computation of the electron density. The computational complexity of this step scales as $\order(N^3)$. Hence, the approach developed in Motamarri et~al.~\cite{motam2013}, subsequently referred to as the Chebyshev filtered subspace iteration for the finite-element basis (ChFSI-FE), scales as $\order(N^3)$ asymptotically (as $M\propto N$), thus limiting accessible systems to a few thousand atoms. In order to address this significant limitation, we build on our prior work~\cite{motam2013} to develop a subspace projection technique that reduces the computational complexity of solving the Kohn-Sham problem. The proposed approach treats both insulating and metallic systems under a single framework, and is applicable for both pseudopotential and all-electron calculations. The key ideas involved in the method for a single self-consistent field (SCF) iteration are discussed below.
\vspace{-0.15in}
\subsection{Chebyshev filtered subspace iteration}\label{Sec:ChebyshevFilter}
\vspace{-0.15in}
Chebyshev filtered subspace iteration (ChFSI)~\cite{saadbook} belongs to the class of subspace iteration techniques which are generalizations of the power method applied to a subspace. As the ground-state electron density, and subsequently the ground-state energy, depends solely on the occupied eigenspace---the vector space spanned by the eigenfunctions corresponding to the occupied states---the ChFSI technique exploits the fast growth property of Chebyshev polynomial in $(-\infty, -1)$ to magnify the relevant spectrum and thereby providing an efficient approach for the solution of the Kohn-Sham eigenvalue problem. We refer to Zhou et al. ~\cite{saad2006} and Motamarri et al.~\cite{motam2013} for the application of this technique to electronic structure calculations in the context of finite-difference  and finite-element discretizations, respectively. We also refer to Garcia et al.~\cite{carlos} for a linear-scaling subspace iteration technique based on Chebyshev filtering for insulating systems.
In the present work, we also adopt the Chebyshev filtering procedure to find the occupied eigenspace at any given SCF iteration. We start with an initial subspace $\overline{\mathbb{V}}^{N}$ of dimension $N$ ($N > N_e/2$) spanned by the set of localized wavefunctions $\{\psi_1(\bx),\psi_2(\bx),\cdots,\psi_N(\bx)\}$ obtained from the previous SCF iteration (cf. section~\ref{Sec:Localization} for details on the construction of the localized wavefunctions). We note that, here, and subsequently, all the electronic fields (wavefunctions and electron density) denote finite-element discretized fields, and we have dropped the superscript $h$ for notational simplicity. We denote by $\btX$ the matrix whose column vectors are the coefficients of expansion of these localized wavefunctions in the L{\"o}wdin orthonormalized spectral finite-element basis ($q_j: 1 \leq j \leq M$). The Chebyshev filtered subspace iteration then proceeds as follows:
\paragraph{Construction of shifted and scaled Hamiltonian:}
 The discretized Hamiltonian $\btH$ is scaled and shifted to construct  $\bar{\bH}$ such that the unwanted (unoccupied) spectrum of $\btH$ is mapped to $[-1,1]$ and the wanted (occupied) spectrum into $(-\infty,-1)$.  Hence
\begin{equation}
\bar{\bH} = \frac{1}{e} (\btH - c \bI)\;\;\text{where}\;\; e = \frac{b - a}{2}\;\;\;\;c = \frac{a+b}{2}\,.
\end{equation}
Here $a$ and $b$ denote the upper bound of the wanted and unwanted spectrum of $\btH$, respectively. The upper bound $b$ is obtained inexpensively using a very small number of Lanczos iterations~\cite{lanczos1950} whose computational complexity scales as $\order (M)$. The upper bound of the wanted spectrum is chosen as the largest Rayleigh quotient of $\btH$ in the occupied eigenspace computed in the previous SCF iteration.

\paragraph{Construction of Chebyshev filter:}
In a given SCF iteration, the action of a Chebyshev filter on $\btX $ is given by
\begin{equation}\label{filter}
\btY = T_m(\bar{\bH}) \btX\,,
\end{equation}
where $\btY$ denotes the matrix whose column vectors are the coefficients of expansion of the Chebyshev filtered wavefunctions $\{\psi^f_1(\bx),\psi^f_2(\bx),\cdots,\psi^f_N(\bx)\}$ expressed in the L{\"o}wdin orthonormalized finite-element basis. We note that the subspace $\mathbb{V}^{N}$ spanned by these Chebyshev filtered wavefunctions is a close approximation to the occupied eigenspace. In the above, the filter $T_m(\bar{\bH})$ is constructed using a Chebyshev polynomial $T_m(x)$ of degree $m$, and the action of the filter on $\btX $ can be recursively computed~\cite{rivlin73} as
\begin{equation}\label{recur}
T_{m}(\bar{\bH})\btX   = \left[2\bar{\bH}\,T_{m-1}(\bar{\bH})  - T_{m-2}(\bar{\bH})\right]\btX \,,
\end{equation}
with $T_0(\bar{\bH})\btX=\btX$ and $T_1(\bar{\bH})\btX=\bar{\bH}\btX$. It is evident from equation~\eqref{recur} that the application of Chebyshev filter on $\btX$ involves matrix-vector multiplications between the discretized Hamiltonian $\btH$ and the vectors obtained during the course of recursive iteration~\eqref{recur}. Further, we note that the finite-element basis functions $N_j: 1 \leq j \leq M$ are local. Hence, the discretized Hamiltonian $\btH$ expressed in the L{\"o}wdin orthonormalized spectral finite-element basis  $q_j: 1 \leq j \leq M$ is sparse, and has the exact same sparsity as $\bH$ if the matrix $\bM$ is evaluated using the GLL quadrature. Thus, if the vectors obtained during the process of recursive iteration of Chebyshev filtering procedure are sparse, the computational complexity of the relevant matrix-vector multiplications scales as $\order(M)$.

In practice, we exploit the locality of finite-element basis to construct the elemental-matrices corresponding to $\bH^{\text{loc}}$, i.e. the individual contributions from each finite-element to $\bH^{\text{loc}}$, without explicitly assembling the global matrix. We note that building these elemental-matrices scales as $\order (M)$. In the case of pseudopotential calculations, we compute $C^{J}_{lm,j}$ at the finite-element level only for those elements in the compact support of $ \Delta V^{J}_{l}$. Further, the vectors obtained during the course of the recursive iteration~\eqref{recur} are truncated using a truncation tolerance $\delta_c$, below which the values are set to zero. The matrix-vector multiplications in the Chebyshev filtering procedure are performed at the finite-element level only if the vector has a non-zero value in the finite-element considered. This leads to the computation of only the non-trivial elemental matrix-vector products, thereby rendering the computational complexity of the Chebyshev filtering procedure to $\order (M)$. The resulting element-level vectors are then assembled to form the global vectors $\btY$. We choose $\delta_c$ to be in the range of $10^{-4}$ to $10^{-14}$ in our subsequent numerical simulations (cf. section~\ref{sec:results}) with looser tolerances being employed in the intial SCF iterations, where the solution is far away from the ground-state solution, and, adaptively employing tighter tolerances in later iterations as the SCF approaches convergence.

We next discuss the localization procedure employed to construct the localized wavefunctions spanning the Chebyshev filtered subspace.
\vspace{-0.15in}
\subsection{Localization and truncation}\label{Sec:Localization}
\vspace{-0.15in}
Various procedures have been employed in the literature to achieve linear-scaling (cf.~\cite{bow2012,godecker99} for a comprehensive review) for pseudopotential calculations, and the use of localized functions spanning the occupied eigenspace (or an approximation to this space) has been one of the ideas exploited in developing linear-scaling algorithms for materials systems with a band-gap. In this regard, we remark that Wannier functions~\cite{wan37,kohn59,kohn73,kohn93} have played an important role. In particular, the maximally localized Wannier functions~\cite{marz97} have been effectively used as an orthogonal localized basis for Kohn-Sham DFT calculations, specifically in periodic systems. However, techniques employing non-orthogonal localized functions have also been proposed~\cite{and68,mauri93,kim95,orde93}, which have better localization properties than orthogonal functions. A discussion on general localization properties of bases spanning the eigenspace is provided in E et al.~\cite{pnas2010} and Ozolins et al.~\cite{pnas2013}. In the present work, we adopt the technique proposed in Garcia-Cervera et al.~\cite{carlos} to construct a localized basis for the subspace $\mathbb{V}^{N}$ spanned by Chebyshev filtered wavefunctions $\{\psi^f_1(\bx),\psi^f_2(\bx),\cdots,\psi^f_N(\bx)\}$. Here, the localized basis is obtained as
\begin{equation}\label{loc}
\text{arg}\;\; \min_{\psi \in \mathbb{V}^{N}, ||\psi|| = 1} \int w(\bx) |\psi(\bx)|^2 \dx\,,
\end{equation}
where $w(\bx)  \geq  0$ is chosen to be a smooth weighting function  $|\bx - \bb_I|^{2p}$, with $p$ being a positive integer and $\bb_I$ denoting a localization center. Such a choice of $w(\bx)$ minimizes the spread of the wavefunctions from a localization center, similar in spirit to the construction of maximally localized Wannier functions~\cite{marz97}. In the present work, we choose $p$ to be 1 and  $\bb_I$ to be an atom center $\bR_I$. Let $n_I$ denote the number of localized functions we desire to compute at every atom center $\bR_I$.
Also, letting
\begin{equation}\label{lincomb}
\psi(\bx) = \sum_{i} \alpha_i \psi^{f}_{i}(\bx) \;\; \in \;\;  \mathbb{V}^{N}\,,
\end{equation}
the minimization problem~\eqref{loc} is equivalent to solving the following generalized eigenvalue problem for the smallest $n_I$ eigenvalues
\begin{equation}\label{loceig}
\bW ^{I}\balpha = \lambda \bS \balpha\,,
\end{equation}
where
\begin{subequations}
\begin{align}\label{wmat}
W^{I}_{ij} &= \int |\bx - \bR_I |^2 \,\psi^{f}_{i}(\bx)\,\psi^{f}_{j}(\bx) \dx\;\;\; i,j = 1\cdots N  \\
S_{ij} &= \int \psi^{f}_{i}(\bx)\,\psi^{f}_{j}(\bx)  \dx\,.
\end{align}
\end{subequations}
In the present work, we choose $n_I$ to be equal to the number of occupied single-atom orbitals corresponding to the $I^{th}$ atom. If $\sum_{I} n_I<N$, then we randomly pick some atoms to compute additional localized functions. We note that we can rewrite $\bW^{I}$ in~\eqref{wmat} using matrix notation to be
\begin{equation}
\bW^{I} = \bL^{T} \bK^{I}\bL\,,
\end{equation}
where columns of the matrix $\bL$ correspond to the nodal values of the wavefunctions $\{\psi^f_1(\bx),\psi^f_2(\bx),\cdots,\psi^f_N(\bx)\}$ obtained after Chebyshev filtering, and
\begin{equation}
K_{ij}^{I} = \int |\bx - \bR_I |^2 \, N_{i}(\bx) N_{j}(\bx) \dx\,.
\end{equation}
We note that the computational complexity of computing the matrix $\bW^{I}$ for an atom $I$ scales as $\order (MN^{2})$ as the wavefunctions $\{\psi^f_1,\psi^f_2,\cdots,\psi^f_N\}$ obtained after Chebyshev filtering need not be sparse. Hence the total computational complexity for $N_a$ atoms ($N_a \propto N$) in a given materials system scales as $\order (MN^3)$. We propose the following procedure in order to reduce the computational complexity to $\order (M)$. We first expand $\bK^I$ in terms of atom $I=0$ as follows:
\begin{align*}
K_{ij}^{I} =& \int |\bx - \bR_0 + \bR_0 - \bR_I|^2\; N_{i}(\bx) N_j(\bx) \dx\\
= & \int \Bigl[|\bx - \bR_0|^{2} \,N_{i}(\bx) N_{j}(\bx) + |\bR_0 - \bR_I|^2 \,N_{i}(\bx) N_{j}(\bx) \\
& + 2 \,(\bx - \bR_0)\,.\, (\bR_0 - \bR_I) \,N_{i}(\bx) N_{j}(\bx) \Bigr] \dx\\
=\,\, & K_{ij}^{0} + |\bR_0 - \bR_I|^2\,M_{ij} + 2\, (R_{0x} - R_{Ix})B_{ij}^{x} \\
& + 2\, (R_{0y} - R_{Iy})B_{ij}^{y} + 2\, (R_{0z} - R_{Iz})B_{ij}^{z}
\end{align*}
where
\begin{subequations}\label{lockmat}
\begin{align}
K_{ij}^{0} &= \int |\bx - \bR_0 |^2 \, N_{i}(\bx) N_{j}(\bx) \dx \,, \\
M_{ij} &= \int  N_{i}(\bx) N_{j}(\bx) \dx \,, \\
B^{x}_{ij} &= \int  (x - R_{0x}) \, N_{i}(\bx) N_{j}(\bx) \dx \,, \\
B^{y}_{ij} &= \int  (y - R_{0y}) \, N_{i}(\bx) N_{j}(\bx) \dx \,, \\
B^{z}_{ij} &= \int  (z - R_{0z}) \, N_{i}(\bx) N_{j}(\bx) \dx \,.
\end{align}
\end{subequations}
We use Gauss-Lobatto-Legendre quadrature rules to evaluate each of the above integrals which renders the matrices in ~\eqref{lockmat} diagonal. Further, the matrix $\bW^{I}$ for any atom $I$ is constructed as a linear combination of five matrices, and is given by
\begin{align}\label{locwmat}
\bW^{I} &= \bL^{T} \bK^{0} \bL + |\bR_0 - \bR_I|^2 \bL^{T} \bM \bL + \nonumber \\
& 2\, (R_{0x} - R_{Ix})  \bL^{T} \bB^{x} \bL  + 2\, (R_{0y} - R_{Iy}) \bL^{T} \bB^{y} \bL + \nonumber \\
& 2\, (R_{0z} - R_{Iz}) \bL^{T} \bB^{z} \bL\,.
\end{align}
We note that the five matrices $\bL^{T} \bK^{0} \bL $, $\bL^{T} \bM \bL $, $\bL^{T} \bB^{x} \bL $, $\bL^{T} \bB^{y} \bL $ and $\bL^{T} \bB^{z} \bL $ are independent of $I$, and can be computed \emph{a priori}. Further, in order to reduce the computational complexity of computing these five matrices, we introduce a truncation tolerance $\delta_w>\delta_c$ ($\delta_w\sim 10^{-4}-10^{-8}$) to truncate the Chebyshev filtered wavefunctions used in the construction of matrices $\bL$ and $\bS$. We note that this truncation in the Chebyshev filtered wavefunctions is introduced only in the construction of $\bL$ and $\bS$, and not in other operations involving the Chebyshev filtered wavefunctions, in particular, the linear combination in~\eqref{lincomb}. Introducing the truncation $\delta_w$ renders a sparse structure to both $\bL$ and $\bS$, and thus the computation of $\bW^{I}$ for all the atoms $I = 1 \cdots N_a $ scales as $\order (M)$. We note that the use of GLL quadrature rules in the evaluation of matrix elements in ~\eqref{lockmat}, as well as, the use of truncation tolerance $\delta_w$ in the construction of $\bL$ and $\bS$ introduces approximation errors in the solution of the eigenvectors $\balpha$ in the generalized eigenvalue problem~\eqref{loceig}. However, we note that these approximations errors do not alter the space $\mathbb{V}^{N}$ spanned by the localized wavefunctions (cf. equation~\eqref{lincomb}), as the vector space remains invariant under any linear combination.

Using the eigenvectors $\balpha$ from the solution of the eigenvalue problem in~\eqref{loceig} for each atom $I$, and the linear combination in~\eqref{lincomb}, the non-orthogonal localized wavefunctions are computed and are denoted by $\{\phi^L_1(\bx),\phi^L_2(\bx),\cdots,\phi^L_N(\bx)\}$ which span $\mathbb{V}^{N}$. In order to provide a compact support for these non-orthogonal localized wavefunctions, we introduce a truncation tolerance $\delta_l$, where the nodal values of these functions that are below this tolerance are set to zero. We note that, upon truncating these wavefunctions, the space spanned by these functions is only an approximation to $\mathbb{V}^{N}$, where the approximation error in the subspace depends on the choice of $\delta_l$. As in the case of $\delta_c$, the truncation tolerance introduced during the Chebyshev filtering, $\delta_l$ is also chosen adaptively with looser tolerances in the initial SCF iterations and progressively employing tighter tolerances as the SCF approaches convergence (see section~\ref{sec:results} for details). We denote by $\bPhi_{L}$ the matrix whose column vectors are the expansion coefficients of these compactly-supported non-orthogonal localized wavefunctions expressed in the L{\"o}wdin orthonormalized finite-element basis. We note that the locality of the non-orthogonal wavefunctions $\phi^L_{i}(\bx): {1 \leq i \leq N}$ renders sparsity to the matrix $\bPhi_{L}$.
\vspace{-0.15in}
\subsection{Subspace projection in the non-orthogonal basis}
\vspace{-0.15in}
We now discuss the steps involved in the subspace projection of the Hamiltonian into the non-orthogonal localized basis represented by $\bPhi_L$.
\setcounter{paragraph}{0}
\vspace{0.1in}
\paragraph{Computation of overlap matrix:}
The overlap matrix $\bS$ resulting from the non-orthogonal localized wavefunctions is given by
\begin{equation}
\bS = \bPhi^{T}_{L} \bPhi_{L}\,.
\end{equation}
We note that the computational complexity of evaluating $\bS$ scales as $\order (N)$ if $\bPhi_{L}$ is a sparse matrix.
\vspace{0.1in}
\paragraph{Computation of projected Hamiltonian:}
We recall from the discussion in section ~\ref{sec:math} that the projection of the Hamiltonian into a non-orthogonal localized basis is given by
\begin{equation}\label{projHamalpA}
\bH^{\phi} =  \bS^{-1} \bPhi^{T}_{L} \btH \bPhi_{L}\,.
\end{equation}
The above equation involves the inverse of the overlap matrix  $\bS^{-1}$, which can be evaluated using scaled third-order Newton-Schulz iteration ~\cite{jansik2007,carlos}. If $\bS$ and $\bS^{-1}$are exponentially localized, $\bS^{-1}$ can be computed in $\order (N)$ complexity~\cite{rubenson2005}. Since the discretized Hamiltonian $\btH$ is sparse, and if the matrix $\bPhi_{L}$ is sparse with a bandwidth independent of $N$, $\bH^{\phi}$ can be computed in $\order (N)$ complexity.

\vspace{-0.15in}
\subsection{Electron density computation}
\vspace{-0.15in}
The Fermi-operator expansion techniques~\cite{godecker99,bow2012} have been widely adopted to avoid explicit diagonalization of the discretized Hamiltonian in order to compute the electron density. One of the widely used Fermi-operator expansion technique~\cite{godecker94,godecker95}, which works for both insulating as well as metallic systems, approximates the Fermi distribution (cf. equation~\eqref{fermidirac1}) by means of a Chebyshev polynomial expansion. The accuracy of such an expansion depends on the smearing parameter in the Fermi distribution, $\sigma$, and the width of the eigenspectrum (spectral width) of discretized Hamiltonian denoted by $\Delta E$. In particular, the degree of polynomial required to achieve a desired accuracy in the approximation~\cite{baer97} of the Fermi distribution is $\order(\frac{\Delta E}{\sigma})$, and is $\order(\sqrt{\Delta E})$ for a given $\sigma$ and the occupied spectrum~\cite{phanish2013b}. We also note that numerous recent efforts have focused on developing alternate approximations to the Fermi distribution~\cite{ozaki2007,ceriotti2008,ozaki2010,linlin2009,linlin2013a} with reduced computational complexity.

One of the major challenges in employing the Fermi-operator expansion technique on a finite-element discretized Hamiltonian is the large spectral width of the Hamiltonian---$\order (10^3)~Ha$ for pseudopotential calculations and  $\order (10^5)~Ha$ for all-electron calculations~\cite{motam2013}---that deteriorates the accuracy of the Fermi-operator expansion. To circumvent this, we employ the Fermi-operator expansion in terms of the subspace projected Hamiltonian $\bH^{\phi}$ whose spectral width is commensurate with that of the occupied eigenspectrum. We recall from equation~\eqref{elecden} that the electron density in terms of the density matrix is given by
\begin{equation}
\rho(\bx) = 2\;\sum_{i,j = 1}^{M} \Gamma_{ij} q_i(\bx) q_j(\bx)\,.
\end{equation}
Using  equations ~\eqref{orthfem} and ~\eqref{denmat} in the above equation, we have
 \begin{align}\label{electrondensity}
 \rho(\bx) &=  2\;\bN^{T}(\bx)\; \bM^{-1/2}\; \Phi_L \;f(\bH^{\phi}) \;\Phi_{L}^{\dag}\; \bM^{-1/2}\bN(\bx) \nonumber \\
&=  2\;\bN^{T}(\bx)\; \bM^{-1/2}\; \Phi_L \;f(\bH^{\phi}) \;\bS^{-1}\;\Phi_{L}^{T} \bM^{-1/2}\bN(\bx)\,,
 \end{align}
where $\bN^{T}(\bx) = [N_1(\bx)\,, N_2(\bx)\,, N_3(\bx),\, \cdots\,, N_M(\bx)] $
and
\begin{equation}\label{fermidirac}
f(\bH^{\phi}) = \frac{1}{1 + \exp\left(\frac{\bH^{\phi} - \mu}{\sigma}\right)}
\end{equation}
with $\mu$ being the chemical potential and $\sigma = k_B\,T$. We note that as the self-consistent field iteration converges, $f(\bH^{\phi})$ represents the finite-temperature density matrix expressed in the non-orthogonal localized basis. We remark that, for a jellium approximation (a simplified representation for a metallic system), it was shown that the finite-temperature density operator exhibits an exponential decay in real-space~\cite{godecker98,arias99}. However, this remains an open question beyond the jellium approximation for metallic systems. For an insulating system with a band-gap, the density operator, even at zero temperature, has an exponential decay in real-space~\cite{godecker99}. Assuming that the density operator decays in real-space, and recalling that $\bPhi_{L}$ represents a localized basis, we note that $f(\bH^{\phi})$ will have a sparse structure with the extent of sparsity depending on the decay properties of the density operator.

We use Chebyshev polynomial expansion to approximate $f(\bH^{\phi})$, and compute the electron density. To this end, we begin by scaling and shifting $\bH^{\phi}$ to obtain $\bH^{\phi}_s$ such that its spectrum lies in $[-1,1]$ and then $f(\bH^{\phi})$ is approximated using a finite number of Chebyshev polynomials~\cite{baer97} as
\begin{equation}\label{chebexp}
 f(\bH^{\phi}) = \sum_{n=0}^{R} a_n(\sigma_s,\mu_{s}) T_n(\bH^{\phi}_s),
 \end{equation}
 where
 \begin{equation}
\bH^{\phi}_s = \frac{\bH^{\phi} - \bar{\epsilon}}{\Delta \epsilon} \;\;\;;\;\;\sigma_s = \frac{\Delta \epsilon}{\sigma} \;\;;\;\; \mu_{s} = \frac{\mu - \bar{\epsilon} }{\Delta \epsilon},
 \end{equation}
 \begin{equation}
\Delta \epsilon = \frac{\epsilon_{max} - \epsilon_{min}}{2}\;\;\;\;\;;\;\;\;\;\;\bar{\epsilon} = \frac{\epsilon_{max} + \epsilon_{min}}{2},
 \end{equation}
and
\begin{equation}
  a_n(\sigma_s,\mu_{s})  = \frac{2 - \delta_{n\,0}}{\pi}\int_{-1}^{1} \frac{T_n(x)}{\sqrt{1-x^2}}\,\frac{1}{1 + e^{\sigma_s(x - \mu_{s})}} dx\,,
  \end{equation}
where $\delta_{ij}$ denotes the Kronecker delta. In the above, $\epsilon_{max}$ and $\epsilon_{min}$ denote the upper and lower bounds for the spectrum of $\bH^{\phi}$. Estimates for $\epsilon_{max}$ and $\epsilon_{min}$ are computed using the Krylov-Schur method~\cite{kschur2001}. As $\bH^{\phi}$ is the projection of the Hamiltonian into a localized basis, $\bH^{\phi}$ is a sparse matrix, and these estimates for the spectral width can be computed in $\order (N)$ complexity. We also remark that if $\bH^{\phi}$ is sufficiently sparse, $f(\bH^{\phi})$ can be computed in $\order(N)$ complexity~\cite{baer97}. Further, we note that the degree $R$ of the Chebyshev expansion in equation~\eqref{chebexp} is proportional to the spectral width $\Delta E = \epsilon_{max}- \epsilon_{min}$ of $\bH^{\phi}$. As discussed earlier, since $\bH^{\phi}$ is the projected Hamiltonian in the space containing the occupied eigenstates and only a few unoccupied eigenstates, the spectral width of $\bH^{\phi}$ is $\order (1\,Ha)$ for pseudopotential calculations and $\order (10\,Ha)$ for all-electron calculations for low atomic numbers. Thus, the Fermi-operator expansion can be computed efficiently and accurately using a Chebyshev polynomial expansion of $\order (100)$ for pseudopotential calculations and $\order (1000)$ for all-electron calculations for moderate temperatures ($\sim500 K$) used in the smearing parameter.

The Fermi-energy ($\mu$), which is required in the computation of the Fermi-operator expansion for $f(\bH^{\phi})$ and the electron density, is evaluated using the constraint in~\eqref{cons1}
\begin{equation}
2\;\text{tr}\left(f(\bH^{\phi})\right) = N_e\,,
\end{equation}
where $N_e$ is the number of electrons in the given system. The above equation is solved using the Newton-Raphson method~\cite{atkinson89} and an initial guess to the Newton-Raphson method is computed using the bisection algorithm~\cite{atkinson89}. It is evident from equation~\eqref{chebexp} that the dependence of the expansion on the Fermi-energy is only in the coefficients of the expansion. Exploiting this fact, the Fermi-energy can be efficiently computed using methods described in Baer et al.~\cite{baer97}, which scales as $\order(N)$. To this end, the $m^{th}$ column of $f(\bH^{\phi})$ is obtained by the application of the expansion in ~\eqref{chebexp} on a unit column vector  $\bv^m$ containing all zeros except at the $m^{th}$ position. Hence,
\begin{equation}
\left[f(\bH^{\phi})\right]_m = \sum_{n=0}^{R}  a_n(\sigma_s,\mu_{s}) \bv_{n}^{m},
\end{equation}
where $\bv_{n}^{m}=T_n(\bH^{\phi}_s)\bv^m$ denotes the $m^{th}$ column of $T_n(\bH^{\phi}_s)$. We note that $\bv_{n}^{m}$ can be computed efficiently using the Chebyshev polynomial recursion given by
\begin{align*}
v_0^{m} &= v^{m}\\
v_1^{m} &= \bH^{\phi} v^{m} \\
v_{n+1}^{m} &= 2\,\bH^{\phi} v_{n}^{m} -  v_{n-1}^{m}\,.
\end{align*}
As the vectors $\bv_{n}^{m}$ can be precomputed and stored, the evaluation of $\text{tr}\left(f(\bH^{\phi})\right)$ for every trial Fermi-energy results in a trivial computational cost.

Upon evaluating the Fermi energy, the band energy ($E_b$) can also be expressed in terms of  $\bH^{\phi}$ as
\begin{equation}
E_b = 2\;\text{tr} \left( f(\bH^{\phi}) \bH^{\phi}\right)\,.
\end{equation}
Finally, we note that the computational complexity of computing the electron density at a quadrature point $\bx$ is independent of the system size. This can be seen by rewriting the expression~\eqref{electrondensity} as
 \begin{align}\label{electrondensity1}
 \rho(\bx) &=  2\;\br^{T}(\bx) \;f(\bH^{\phi}) \;\bS^{-1}\;\br(\bx)\,,
 \end{align}
where $\br(\bx) = \bPhi_{L}^{T}\bM^{-1/2}\bN(\bx)$ is a sparse vector if $\bPhi_L$ is sparse. As $f(\bH^{\phi})\bS^{-1}$ can be pre-computed \emph{a priori}, the computation of $\rho(\bx)$ is independent of the system size. Since the total number of quadrature points scales linearly with system size, computation of electron density for a given material system is $\order (M)$ complexity.

%% file: results.tex
In this section, we investigate the accuracy, performance, and scaling of the proposed subspace projection technique. As benchmark systems, we consider non-periodic three dimensional systems involving metallic, insulating and semi-conducting materials systems. The benchmark metallic systems chosen for this study include aluminum nano-clusters of varying sizes, containing $3\times 3\times 3$ (172 atoms), $5\times 5\times 5$ (666 atoms), $7\times 7\times 7$ (1688 atoms) and $9 \times 9 \times 9$ (3430 atoms) face-centered-cubic (fcc) unit-cells. The benchmark insulating systems chosen for this study include alkane chains of varying lengths, containing 101, 302, 902, 2702 and 7058 atoms. Silicon nano-clusters containing $1\times 1\times 1$ (252 electrons), $2\times 1\times 1$ (434 electrons), $2\times 2\times 1$ (756 electrons),  $2 \times 2 \times 2$ (1330 electrons) and $3 \times 3 \times 3$ (3920 electrons) diamond-cubic (dia) unit-cells are chosen for the benchmark semi-conducting materials systems. Norm-conserving Troullier-Martins pseudopotentials~\cite{tm91} have been employed in the case of aluminum nano-clusters and alkane chains, whereas all-electron calculations have been performed in the case of silicon nano-clusters. In all our simulations, we use the n-stage Anderson~\cite{andmix65} mixing scheme on the electron density in self-consistent field iteration of the Kohn-Sham problem. Further, all the numerical simulations reported in this section are conducted using a parallel implementation of the code based on MPI, and are executed on a parallel computing cluster with the following specifications: dual-socket eight-core Intel Core Sandybridge CPU nodes with 16 processors (cores) per node, 64 GB memory per node, and 40 Gbps Infiniband networking between all nodes for fast MPI communications.

As discussed in section~\ref{sec:fem}, the linear-scaling of the proposed subspace projection technique relies on the locality of the finite-element basis ($q_{i}:1\leq i\leq M$) as well as the localized wavefunctions ($\phi^L_{j}:1\leq j \leq N$), and subsequently, the sparsity of the various matrices involved in our formulation ($\btH, \bH^{\phi}, \bPhi_{L}, \bS, \bW^{I}$). We recall that the compact support of localized wavefunctions and the sparsity in various matrices is achieved by introducing truncation tolerances $\delta_c$ (in Chebyshev filtering) and $\delta_l$ (for localized wavefunctions). Further, as mentioned in section 4, and elaborated subsequently, the truncation tolerances are chosen adaptively with looser tolerances being employed in the initial SCF iterations and progressively tightening these during the course of the SCF convergence. As demonstrated in our benchmark studies, such an adaptive tolerance provides significant computational efficiency while retaining the accuracy of the solution to the Kohn-Sham DFT problem. We note that the sparsity of various matrices is governed by the eigenspectrum of discrete Hamiltonian as well as the values of the truncation tolerances employed, which changes with each SCF iteration. In the present implementation of the subspace projection technique, we employ efficient parallel data-structures provided by PETSc package~\cite{petsc}, where necessary, to represent various matrices and perform arithmetic operations between them. We observe in our simulations that the operations involving sparse data-structures provided by PETSc are efficient only when the fraction of non-zero entries (density fraction) in the matrix is $<1-2\%$. Beyond this density-fraction, the overhead cost of using a sparse data-structure is prohibitively expensive, and it is more efficient to use dense data-structures. To maximize the computational efficiency, in the present work, we employ sparse data-structures only when the density fraction is $<2\%$, beyond which we switch to dense data-structures. This is the main source of deviation from linear-scaling, in practice, in the present implementation.

In order to assess the accuracy, performance and scaling of the proposed approach, we use as reference the recently developed Chebyshev filtered subspace iteration for the finite-element basis~\cite{motam2013} (ChFSI-FE). The ChFSI-FE involves the projection of the Hamiltonian into an orthogonal basis spanning the Chebyshev filtered space, and an explicit computation of the eigenvalues and eigenvectors of the projected Hamiltonian to estimate the electron density. It was demonstrated~\cite{motam2013} that ChFSI-FE technique with the use of higher-order finite-element discretization presents a computationally efficient real-space approach for Kohn-Sham DFT calculations, which can handle both pseudopotential and all-electron calculations. We also note that the accuracy of ChFSI-FE was ascertained, on benchmark problems of varying sizes, using ABINIT software~\cite{ABINIT} for pseudopotential calculations and GAUSSIAN software~\cite{gaussian} for all-electron calculations. In the present work, we use ChFSI-FE as our reference to assess the accuracy and performance of the proposed approach.

\subsection{Aluminum nano-clusters: Pseudopotential study}~\label{sec:Aluminum_nanoclusters}
We consider aluminum nano-clusters formed from fcc unit cells with lattice spacing of $7.45~a.u$.. The norm-conserving Troullier-Martins pseudopotential in the Kleinman-Bylander form~\cite{tm91,bylander82} is employed in these simulations. We consider the $3s$ and $3p$ components to compute the projector, while the $3d$ component is chosen to be the local component of the pseudopotential. The pseudopotentials are generated using the fhi98pp~\cite{fhi98pp} software and the default cut-off radii are used for $3s$, $3p$ and $3d$ components, which are $1.8~a.u.$, $2.0~a.u.$ and $2.15~a.u.$, respectively. In order to choose finite-element meshes which provide a discretization error of less than $5~meV$ per atom for the various nano-clusters considered in the present study, we first obtain the converged ground-state energy for the aluminum cluster containing $3 \times 3 \times 3$ fcc unit cells which comprises of 172 atoms with 516 electrons. To this end, we use a sequence of increasingly refined fourth-order spectral finite-element (HEX125SPECT) meshes on a cubic simulation domain of side $400~a.u.$ employing Dirichlet boundary conditions. Here, and subsequently, we use the \emph{a priori} mesh adaption techniques developed in Motamarri et.~al~\cite{motam2013} to construct the finite-element meshes, and refer to this prior work for details. For these sequence of meshes, the ground-state energy is computed using ChFSI-FE with a Chebyshev filter of degree 20 and employing a Fermi smearing parameter of $0.00158~Ha$ (T=500K).
The computed discrete ground-state energies ($E_h$) for these meshes are tabulated in Table~\ref{tab:conv3x3x3Cluster}, where $h$ denotes a measure of the finite-element mesh-size. The extrapolation procedure proposed in Motamarri et.~al~\cite{motam2013} allows us to estimate the ground-state energy in the limit as $h\to 0$, denoted by $E_0$. Using this extrapolation procedure, we computed the ground-state energy per atom to be $E_0 = -56.6966935~eV$. In order to ascertain the accuracy of this extrapolated ground-state energy, we conducted a plane wave simulation with ABINIT using a cell-size of $80~a.u.$ and an energy cut-off of $30~Ha$ with one k-point (the most refined calculation possible within our memory limitations). In the case of ABINIT, the ground-state energy per atom was found to be $-56.6966719~eV$, which only differs from $E_0$ by $0.02~meV$.
\begin{table}[htbp]
\caption{\small{Convergence of the finite-element discretization (HEX125SPECT element) for a $3\times 3\times 3$ fcc aluminum cluster.}}
 \begin{center}
 \begin{tabular}{|c|c|c|}
   \hline
 Degrees of & Energy per atom & Relative error  \\
 freedom (DoF) & (eV) & $\big|\frac{E_h - E_0}{E_0}\big|$ \\ \hline\hline
$222,553$ &  -55.5622017  & 2.0 $\times 10^{-2}$   \\ \hline
$1,760,305$  & -56.6677104 & 5.1 $\times 10^{-4}$  \\ \hline
$14,003,809$ & -56.6966331 & 1.06 $\times 10^{-6}$ \\\hline \hline
\end{tabular}
\end{center}\label{tab:conv3x3x3Cluster}
\end{table}

We next choose a finite-element mesh with fifth-order spectral finite-elements (HEX216SPECT) for the $3 \times 3 \times 3$ aluminum nano-cluster, where the mesh-size is chosen such that the discretization error, measured with respect to $E_0$, is less than $5~meV$ per atom. Using the same characteristic mesh-size, which is expected to result in a similar discretization error, we construct the finite-element meshes for the varying sizes of aluminum clusters containing $5\times 5\times 5$ (666 atoms), $7\times 7\times 7$ (1688 atoms) and $9 \times 9 \times 9$ (3430 atoms) fcc unit-cells. Using a Chebyshev filter of degree 20 and a Fermi smearing parameter of $0.00158~Ha$ (T=500K), we compute the reference ground-state energies using ChFSI-FE. These are tabulated in Table~\ref{tab:alumEnergies}.

We next compute the ground-state energies using the proposed subspace projection algorithm (cf. section~\ref{sec:fem}) with identical meshes and parameters (Chebyshev filter degree, Fermi smearing parameter) employed in our reference calculations. The polynomial degree $R$ used in the Chebyshev expansion of the Fermi-function of the projected Hamiltonian~\eqref{chebexp} is chosen to be $400$. We recall that we use truncation tolerances ($\delta_c$, $\delta_l$) to achieve compact support for the non-orthogonal localized wavefunctions ($\bPhi_L$) and sparsity in the various matrices involved in our formulation. In the initial SCF iterations we use looser truncation tolerances, and employ tighter tolerances as the self-consistent iteration proceeds towards convergence. In particular, we choose $\delta_c=\delta_l$ and vary the truncation tolerance as a function of the relative change in the ground-state energy between two successive SCF iterations ($\delta E_r$). The specific choice of the truncation tolerance employed in the present study is
\begin{equation}\label{delta_Al}
\delta_c = \begin{cases}
C_1 & \text{if $\delta E_r \geq 1$},\\
C_1 (\delta E_r)^{p} & \text{if $10^{-1.5} \leq \delta E_r  < 1$},\\
C_2 (\delta E_r)^{q} & \text{otherwise}\,.
\end{cases}
\end{equation}
Here, we use a truncation tolerance of $C_1 = 10^{-4}$ when $\delta E_r$ is greater than 1, then use a power law (cf.~equation~\eqref{delta_Al}), subsequently, with $p = \frac{2}{3}$ and $q = \frac{3}{2}$. In order to ensure continuity in $\delta_c$, $C_2$ is chosen to be $0.00178$. We note that the choice of this truncation tolerance is arbitrary, however, this form of the truncation tolerance has provided robust convergence of the SCF for all benchmark calculations, where the number of SCF iterations using the subspace projection approach is only marginally ($\sim 4-5$ iterations) greater than those using ChFSI-FE. In all our simulations, using the proposed method as well as ChFSI-FE, the SCF is terminated when $\delta E_r < 10^{-7}$ for three successive iterations.

\begin{figure}[htpb]
\vspace{-1.0in}
\begin{center}
\includegraphics[width=0.5\textwidth]{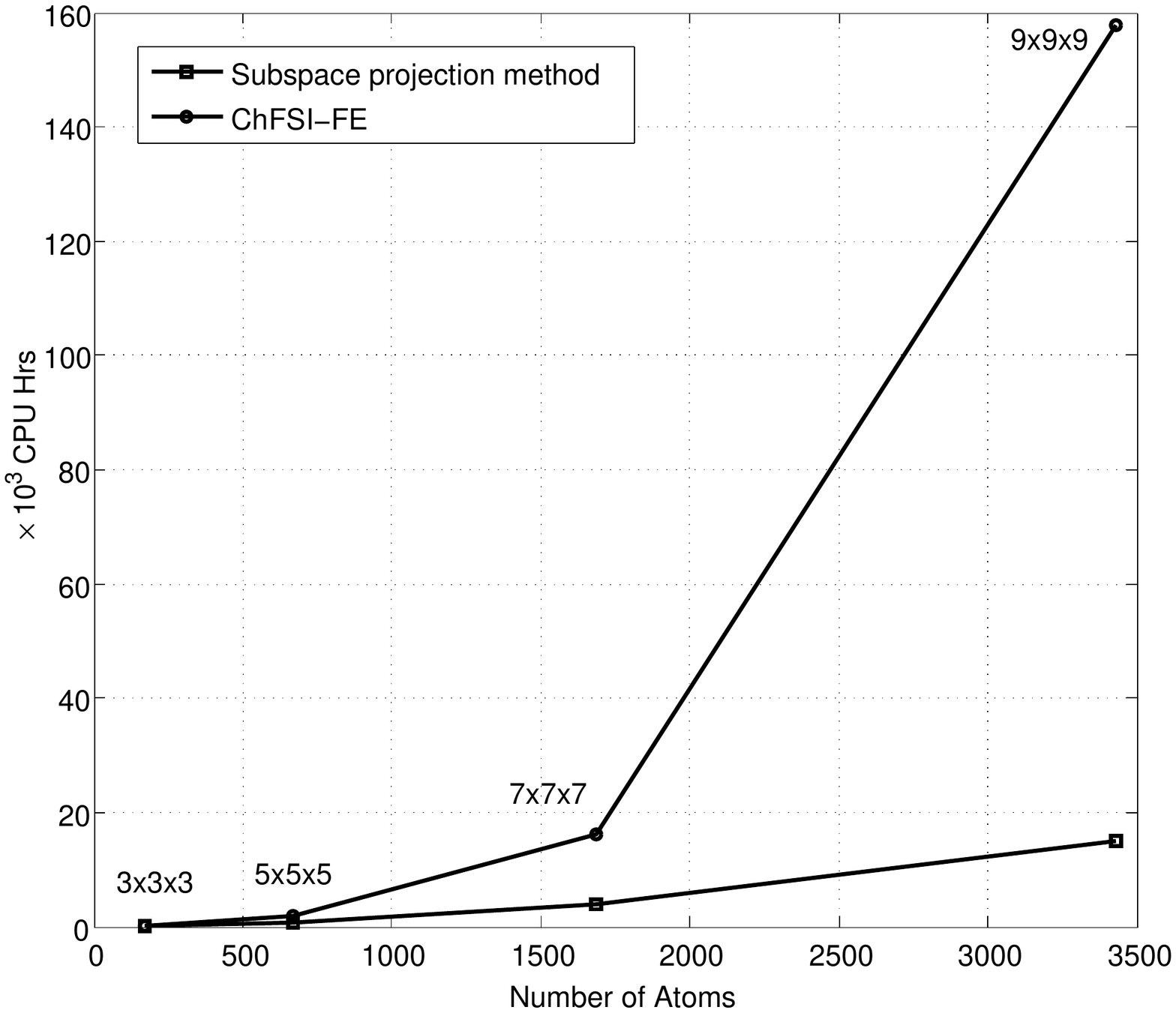}
\vspace{-1.0in}
\caption{\small{Total computational times for the proposed method and ChFSI-FE. Case study: Aluminum nano-clusters.}}
\label{fig:alumScaling}
\end{center}
\vspace{-1.2in}
\begin{center}
\includegraphics[width=0.5\textwidth]{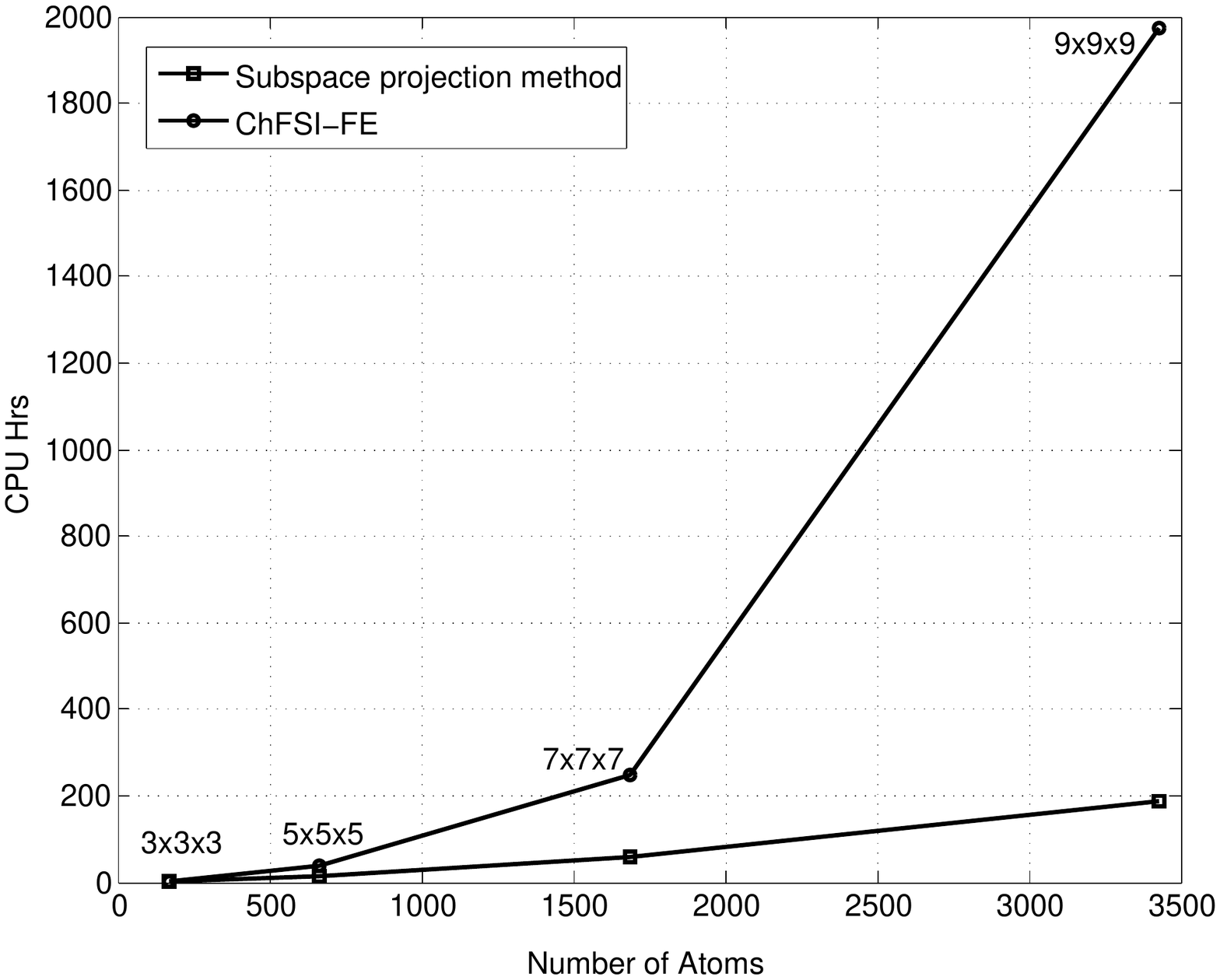}
\vspace{-1.0in}
\caption{\small{Average computational times per SCF iteration for the proposed method and ChFSI-FE. Case study: Aluminum nano-clusters.}}
\label{fig:alumScalingPerSCF}
\end{center}
\begin{center}
\includegraphics[width=0.45\textwidth]{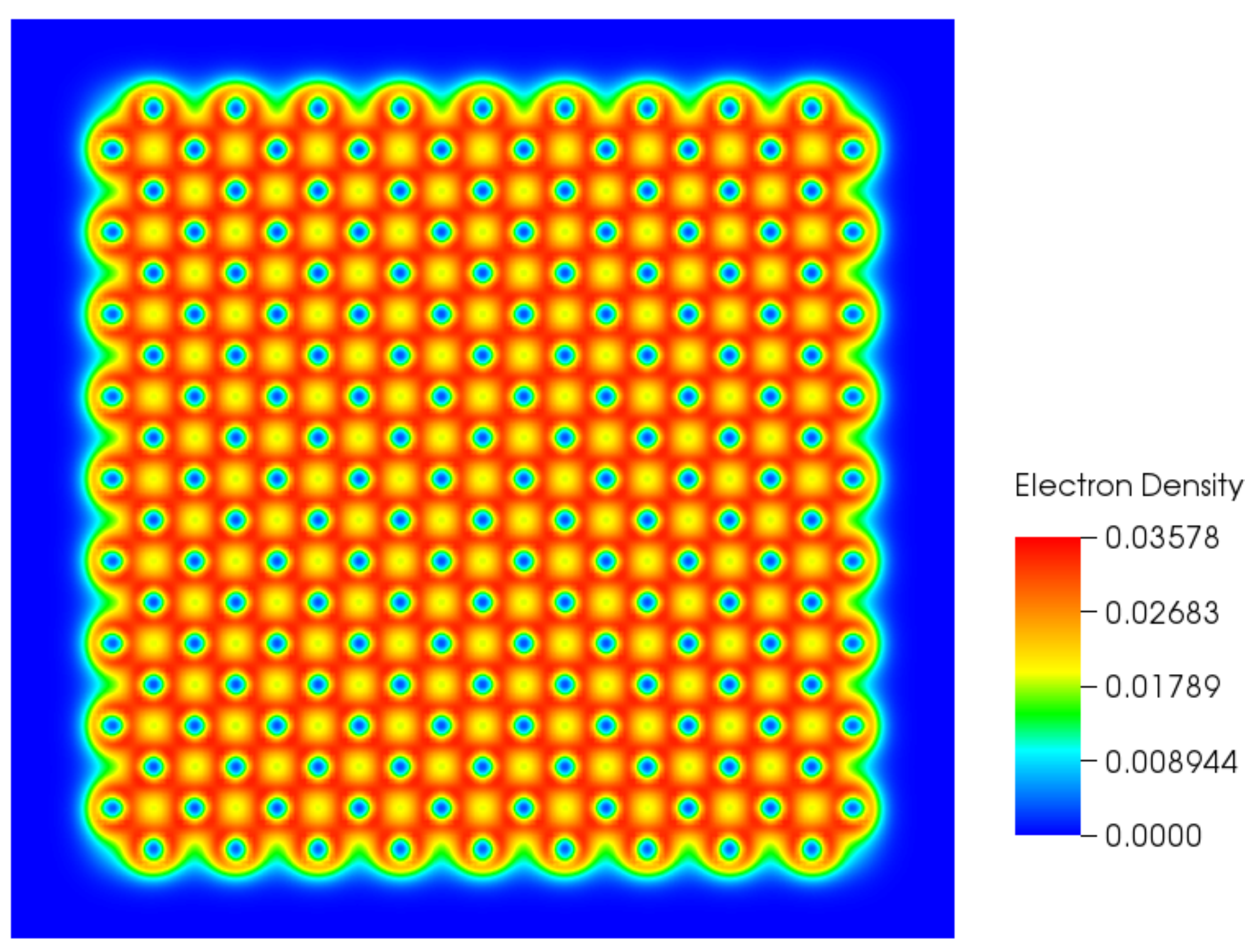}
\caption{\small{Electron density contours on the mid-plane of $9\times 9 \times 9$ fcc aluminum nano-cluster.}}
\label{fig:alumcontour}
\end{center}
\end{figure}


\begin{table}[htbp]
\caption{\small{Ground-state energies per atom (eV) for the various sizes of aluminum nano-clusters computed using the proposed subspace projection algorithm technique and ChFSI-FE~\cite{motam2013}.}}
\label{tab:energies}
 \begin{center}
 \begin{tabular}{|c|c|c|c|}
   \hline
Cluster & DoF & Proposed Method  & ChFSI-FE \\ \hline\hline
3x3x3 &$1,107,471$ &  $-56.694963 $ &  $-56.6949697$ \\ \hline
5x5x5 & $4,363,621$  &   $-56.876491 $ & $-56.876518$  \\ \hline
7x7x7 & $11,085,371$ &  $-56.959021 $ & $-56.9623511$ \\ \hline
9x9x9 &$22,520,721$ &  $-57.010587$ & $-57.0145334$\\ \hline \hline
\end{tabular}
\end{center}\label{tab:alumEnergies}
\end{table}
The ground-state energies computed using the proposed subspace projection algorithm for the range of nano-clusters considered in this work are tabulated in Table~\ref{tab:alumEnergies}. The above results show that the proposed subspace projection technique provides good accuracies in the ground-state energies, where the computed energies are within $5~meV$ per atom of the reference energies computed using ChFSI-FE. The computational times for the full SCF convergence for the range of nano-clusters are plotted in figure~\ref{fig:alumScaling} for the proposed method, as well as, the reference calculations using ChFSI-FE. In this plot, the computational time denoted on the Y-axis is the total computational CPU time in hours (CPU time = Number of cores $\times$ wall-clock time in hours). These results show that the proposed approach is computationally efficient, compared to ChFSI-FE, for system sizes beyond 172 atoms. Further, we note that the subspace projection technique provides a factor of $\sim 8.5$ speedup for the nano-cluster containing 3430 atoms. Using these computational times, we estimated the scaling of the proposed approach and our reference calculations using ChFSI-FE as a function of system size (number of atoms). The scaling for the proposed approach is found to be approximately $\order (N^{1.66})$, while the scaling in the case of ChFSI-FE is $\order (N^{2.37})$. The deviation from linear-scaling of the proposed method, in practice, is primarily due to two factors. Firstly, we observe that the number of SCF iterations increase with increasing system size. For instance, the number of SCF iterations increase from 45 SCF iterations for the $3 \times 3 \times 3$ nano-cluster to 80 iterations for the $9 \times 9 \times 9$ nano-cluster. This issue of increasing SCF iterations with system size can potentially be mitigated using improved mixing schemes~\cite{linlin2013b}. Figure~\ref{fig:alumScalingPerSCF} shows the computational time per SCF iteration, and the scaling with system size is found to be $\order (N^{1.46})$ for the proposed subspace projection method and $\order (N^{2.17})$ for ChFSI-FE. The second factor which results in the deviation of the proposed subspace projection approach from being linear-scaling is due to the use of dense data-structures when the density fraction is above $2\%$. Although the use of dense data-structure affects the overall scaling, it is still computationally efficient in comparison to using sparse data-structures for density fractions greater than $2\%$. We refer to the Appendix for the scaling of various components of the subspace projection algorithm and a discussion. Figure~\ref{fig:alumcontour} shows the electron density contours on the mid-plane of $9\times 9\times 9$ nano-cluster computed using the proposed subspace projection technique.

\subsection{Alkane chains: Pseudopotential study}
\begin{figure*}[t]
\begin{center}
\includegraphics[width=\textwidth]{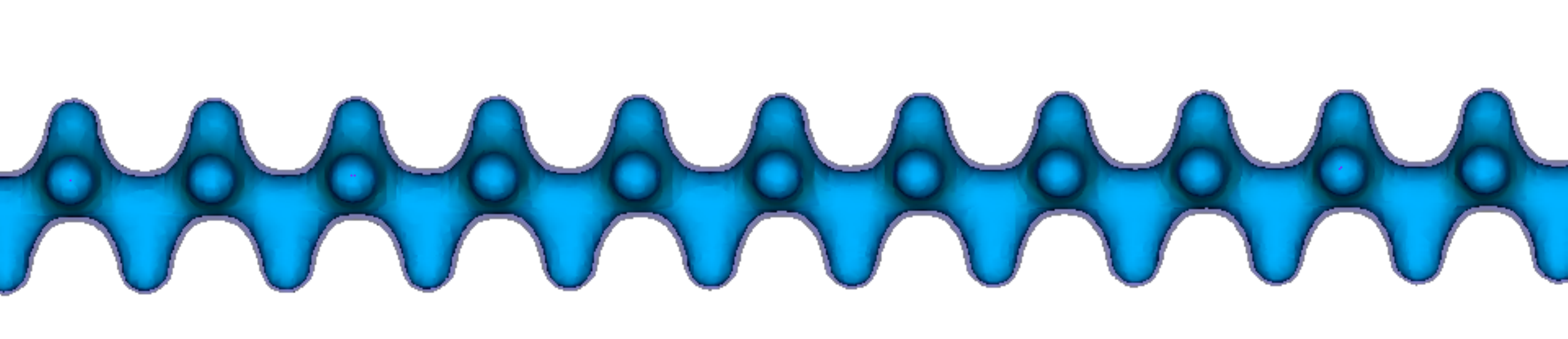}
\caption{\small{Electron density isocontours of $C_{900}H_{1802}$}.}
\label{fig:alkanecontours}
\end{center}
\end{figure*}
We next consider three-dimensional alkane chains with individual repeating units of $\text{C}\text{H}_2$ with C-C and C-H bond lengths to be 2.91018 a.u. and 2.0598 a.u., respectively. The H-C-H and C-C-C bond angles are taken to be 109.47$^{0}$.  Norm-conserving Troullier-Martins pseudopotential in the Kleinman-Bylander form~\cite{tm91,bylander82} has been employed in these simulations. In the case of carbon, we consider the 2s component to compute the projector in the Kleinman-Bylander form, while the 2p component is chosen to be the local component of the pseudopotential. We consider the local pseudopotential for hydrogen corresponding to the 1s component. The pseudopotentials are generated using the software package fhi98PP~\cite{fhi98pp} using the default cut-off radii, which is 1.5 a.u. for both 2s and 2p components of carbon and 1.3 a.u. for the 1s component of hydrogen. In order to choose finite-element meshes with discretization errors less than $5~meV$ per atom for various alkane chains, we first consider the case of $\text{C}_{33}\text{H}_{68}$ containing 101 atoms with 200 valence electrons and obtain the converged value of the ground-state energy. To this end, as in the case of aluminum nano-clusters (cf. section~\ref{sec:Aluminum_nanoclusters}), we use a sequence of increasingly refined fourth-order spectral finite-elements (HEX125SPECT) on a cuboidal simulation domain of dimensions $100~a.u. \times 100~a.u. \times 220~a.u.$. For these sequence of meshes, the ground-state energy $E_h$ is computed using ChFSI-FE with a Chebyshev filter of degree 35 and employing a Fermi smearing parameter of $0.00158~Ha$ (T=500K), and these results are tabulated in Table~\ref{tab:convC33H68}. Using the extrapolation procedure~\cite{motam2013}, we find the reference ground-state energy per atom to be $E_0 = -61.44173873~eV$. Further, we also compared our reference ground-state energy with ABINIT using a cuboidal simulation of size $40~a.u. \times 40~a.u. \times 160~a.u.$ and an energy cut-off of $30~Ha$ with one k-point (the most refined calculation possible within our memory limitations). In the case of ABINIT, the ground-state energy per atom was found to $-61.44219366~eV$, which differs from $E_0$ by $\sim 0.4~meV$.
\begin{table}[htbp]
\caption{\small{Convergence of the finite-element discretization for $\text{C}_{33}\text{H}_{68}$ using HEX125SPECT element.}}
 \begin{center}
 \begin{tabular}{|c|c|c|}
   \hline
 Deg. of freedom & Energy per atom (eV) & Relative error  \\ \hline\hline
$391,893$ & -61.31623538  & 2.0 $\times 10^{-3}$   \\ \hline
$3,096,585$  & -61.43711471 & 7.5 $\times 10^{-5}$  \\ \hline
$24,621,969$ & -61.44173469& 5.01 $\times 10^{-8}$ \\\hline \hline
\end{tabular}
\end{center}\label{tab:convC33H68}
\end{table}

We next choose a finite-element mesh with fifth-order spectral finite-elements (HEX216SPECT) for $\text{C}_{33}\text{H}_{68}$, where the mesh-size is chosen such that the discretization error, measured with respect to $E_0$, is less than $5~meV$ per atom. Using the same characteristic mesh-size which is expected to result in a similar discretization error, we construct finite-element meshes for varying lengths of alkane chains, namely $\text{C}_{100}\text{H}_{202}$ (302 atoms), $\text{C}_{300}\text{H}_{602}$ (902 atoms), $\text{C}_{900}\text{H}_{1802}$ (2702 atoms) and $\text{C}_{2350}\text{H}_{4702}$ (7052 atoms). Using a Chebyshev filter of degree 35 and a Fermi smearing parameter of $0.00158~Ha$ (T=500K), we compute the reference ground-state energies using ChFSI-FE for these systems, and are tabulated in Table~\ref{tab:alkaneEnergies}.

The subspace projection algorithm is then used to compute the ground-state energies  using identical meshes and parameters (Chebyshev filter degree, Fermi smearing parameter) employed in our reference calculations. The polynomial degree $R$ used in the Chebyshev expansion of the Fermi function of the projected Hamiltonian ~\eqref{chebexp} is chosen to be 400. Just as in the case of aluminum nano-clusters, we use adaptive truncation tolerances ($\delta_c$, $\delta_l$). In particular, we choose $\delta_c = \delta_l$ and use the specific choice of truncation tolerance given in equation ~\eqref{delta_Al} with identical values of $C_1$, $p$ and $q$ as employed in the case of our previous benchmark calculations involving aluminum nano-clusters. In all our simulations, using the proposed method as well as ChFSI-FE, the SCF is terminated when $\delta E_r < 10^{-7}$ for three successive iterations.

\begin{table}[htbp]
\caption{\small{Comparison of ground-state energies for various alkane chains.}}
 \begin{center}
 \begin{tabular}{|c|c|c|c|}
   \hline
Alkane Chain & DoF & Proposed Method & ChFSI-FE  \\ \hline\hline
C$_{33}$ H$_{68}$ &$870,656$ &  $-61.438671 $ & $-61.438680$  \\ \hline
C$_{100}$ H$_{202}$ & $2,491,616$  &   $-62.041530 $ & $-62.041532$\\ \hline
C$_{300}$ H$_{602}$ & $7,354,496$ &   $-62.240148 $ & $-62.240277$\\ \hline
C$_{900}$ H$_{1802}$ &$21,943,138$ &  $-62.303101$ & $-62.303608$\\ \hline \hline
\end{tabular}
\end{center}\label{tab:alkaneEnergies}
\end{table}

\begin{figure}[t]
\begin{center}
\includegraphics[width=0.45\textwidth]{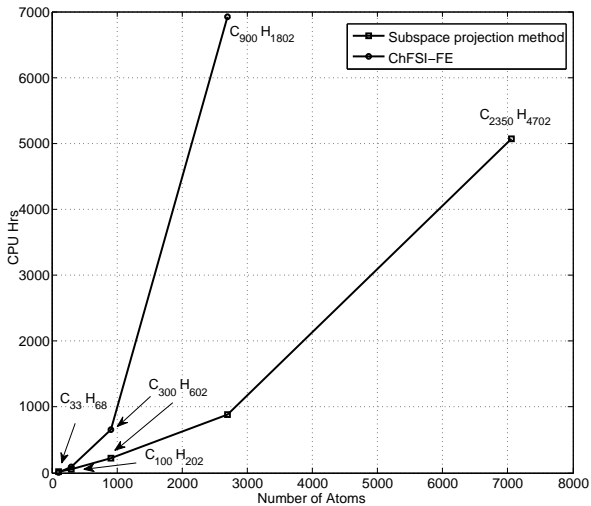}
\vspace{-0.7in}
\caption{\small{Total computational times for the proposed method and ChFSI-FE. Case study: Alkane chains.}}
\label{fig:alkaneScaling}
\end{center}
\end{figure}

\begin{figure}[t]
\begin{center}
\includegraphics[width=0.45\textwidth]{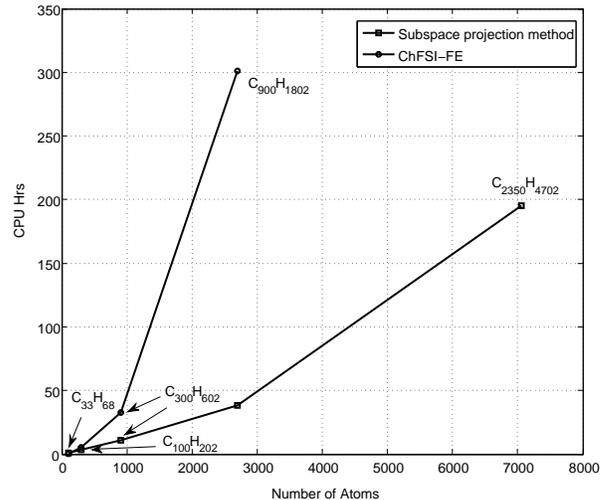}
\vspace{-0.7in}
\caption{\small{Average computational times per SCF iteration for the proposed method and ChFSI-FE. Case study: Alkane chains.}}
\label{fig:alkaneScalingPerSCF}
\end{center}
\end{figure}

The ground-state energies computed using the proposed subspace projection algorithm for different lengths of alkane chains considered in this work are tabulated in Table~\ref{tab:alkaneEnergies}. These results indicate that the proposed method provides good accuracies in the ground-state energies, where the computed energies are within $1~meV$ per atom of the reference energies computed using ChFSI-FE. The computational times for the full SCF convergence for varying lengths of alkane chains are plotted in figure~\ref{fig:alkaneScaling} for the proposed method, as well as, the reference calculations using ChFSI-FE. These results indicate that the proposed approach is computationally efficient, compared to ChFSI-FE, for system sizes beyond 101 atoms and provides a factor of $\sim 8$ speedup for the alkane chain containing 2702 atoms. Using these computational times, the estimated scaling for the proposed approach is found to be approximately $\order (N^{1.33})$, while scaling in the case of ChFSI-FE is $\order (N^{2.13})$. The average computational time per SCF iteration is shown in figure~\ref{fig:alkaneScalingPerSCF}, and the scaling with system size is found to be $\order (N^{1.18})$ for the proposed subspace projection method and $\order (N^{1.98})$ for ChFSI-FE. We refer to the Appendix for the scaling of various components of the subspace projection algorithm and a discussion. We note that the proposed subspace projection technique exhibits better scaling behavior for the alkane chains, which is an insulating system, in comparison to the aluminum nano-clusters, which is a metallic system. This is due to the better localization of the wavefunctions in the insulating system in comparison to the metallic system (cf. Appendix). Figure~\ref{fig:alkanecontours} shows the isocontours of electron density of the alkane chain containing 2702 atoms.



\subsection{Silicon nano-clusters: All-electron study}
We consider silicon nano-clusters comprising of diamond-cubic unit cells with a lattice constant of $10.26~a.u.$ and conduct all-electron calculations to test the performance of the subspace projection method. In order to choose finite-element meshes which provide a discretization error of less than $5~mHa$ per atom for various nano-clusters, we first obtain a converged value of the ground state energy by conducting a very refined simulation using the GAUSSIAN package ~\cite{gaussian} on silicon nano-cluster containing $1 \times 1 \times 1$ unit-cell which comprises of $18$ atoms with $252$ electrons. To this end, we employ the polarization consistent DFT basis sets (pc-n) and introduce them as an external basis set in GAUSSIAN package. Using the most refined pc-4 basis set, the computed ground-state energy per atom ($E_0$) is $-288.3179669~Ha$.

We next chose a finite-element mesh with fifth-order spectral finite-elements (HEX216SPECT) for the $1 \times 1 \times 1$ silicon nano-cluster, where the discretization error, measured with respect to $E_0$, is less than $3~mHa$. Using the same characteristic mesh-size, we construct the finite-element meshes for the varying sizes of silicon nano-clusters containing $1\times 1\times 1$ (252 electrons), $2\times 1\times 1$ (434 electrons), $2\times 2\times 1$ (756 electrons),  $2 \times 2 \times 2$ (1330 electrons) and $3 \times 3 \times 3$ (3920 electrons) diamond-cubic unit cells. The finite-element mesh is locally refined near the nuclei since all-electron calculations involve highly oscillatory wavefunctions near the nuclei. The order of the Chebyshev filter thus required in these simulations is around $1000$ to effectively filter out the unwanted spectrum. Using a Chebyshev filter of degree 1000, and a Fermi smearing parameter of $0.00475~Ha$ (T=1500K), we compute the reference ground-state energies using ChFSI-FE which are tabulated in Table~\ref{tab:siliconEnergies}.

\begin{figure}[htbp]
\vspace{-0.8in}
\begin{center}
\includegraphics[width=0.45\textwidth]{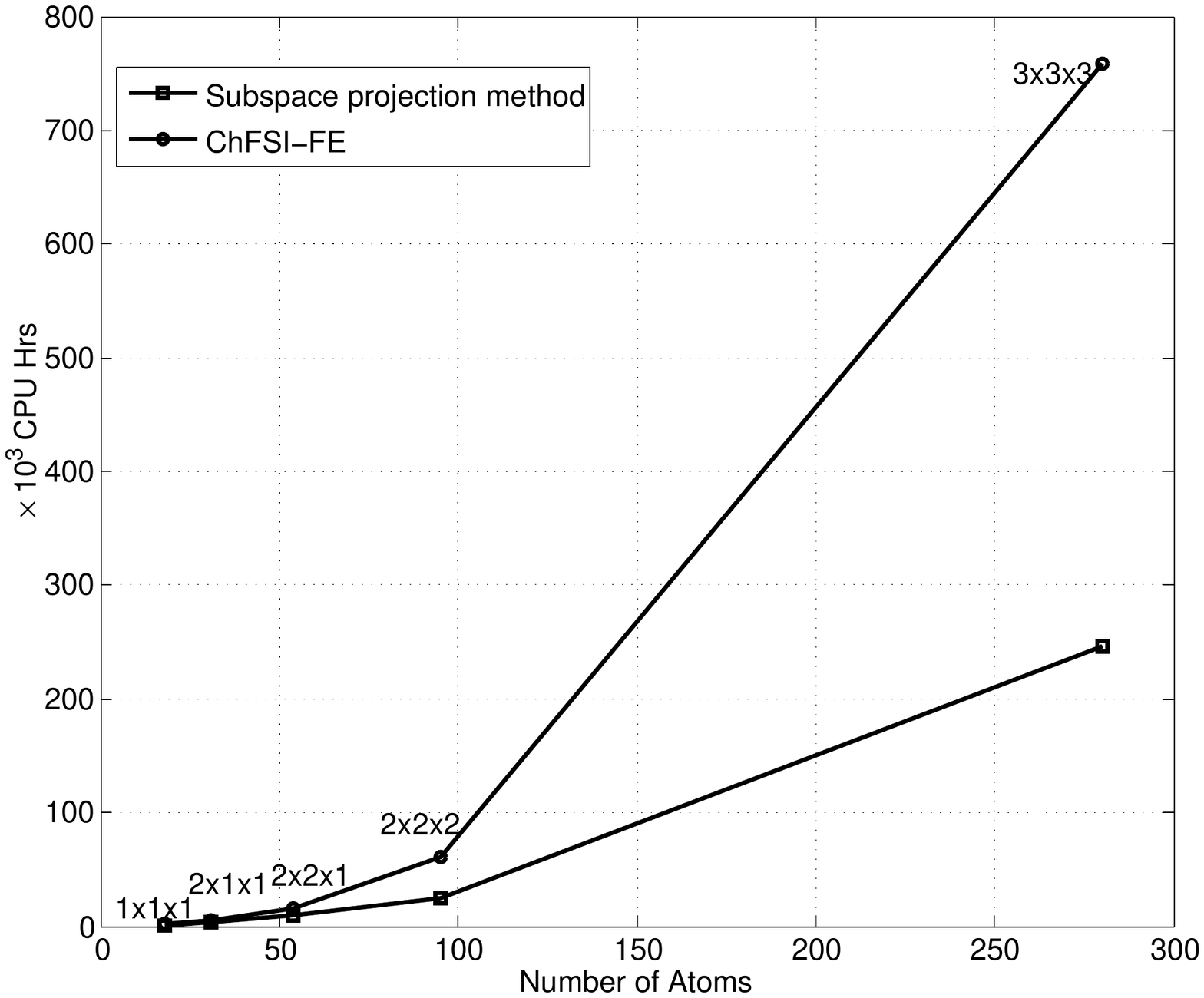}
\vspace{-0.7in}
\caption{\small{Total computational times for the proposed method and ChFSI-FE. Case study: Silicon nano-clusters.}}
\label{fig:siliconScaling}
\end{center}
\vspace{-0.8in}
\begin{center}
\includegraphics[width=0.45\textwidth]{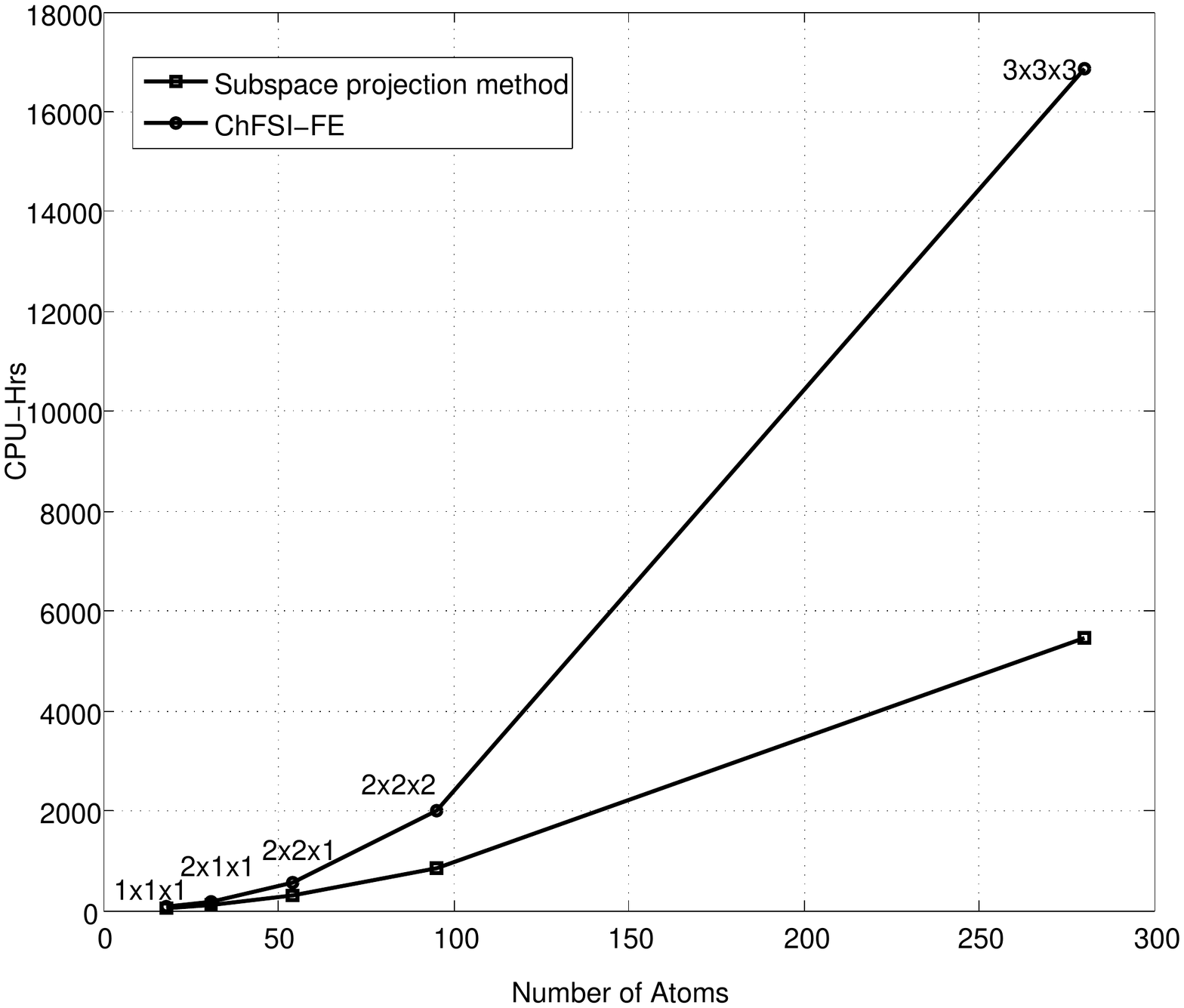}
\vspace{-0.7in}
\caption{\small{Average computational times per SCF iteration for the proposed method and ChFSI-FE. Case study: Silicon nano-clusters.}}
\label{fig:siliconScalingPerSCF}
\end{center}
\begin{center}
\includegraphics[width=0.45\textwidth]{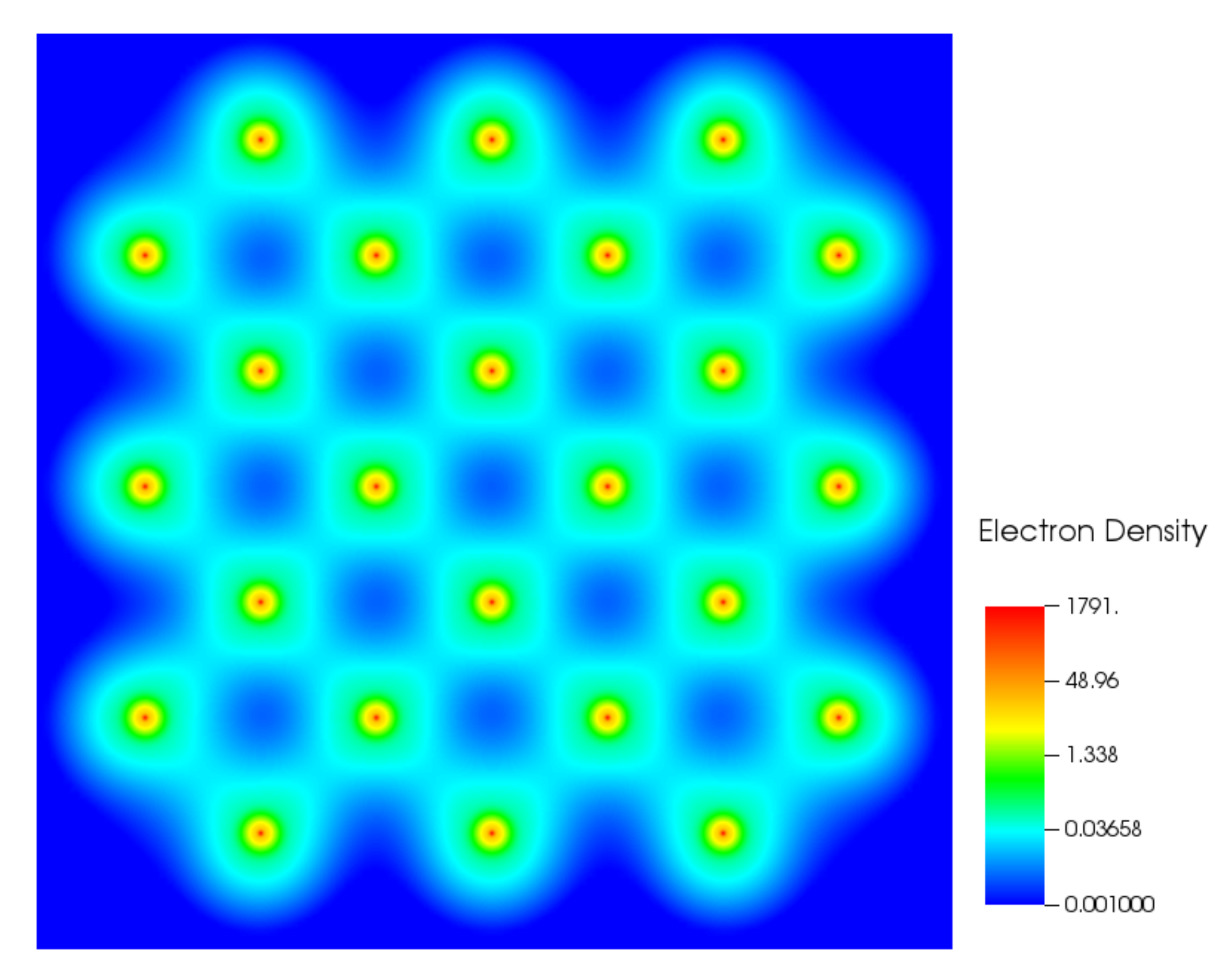}
\caption{\small{Electron density contours on the mid-plane of $3\times 3 \times 3$ silicon nano-cluster.}}
\label{fig:Sicontour}
\end{center}
\end{figure}

The subspace projection algorithm is then used to compute the ground-state energies with identical meshes and parameters employed in our reference calculations. The polynomial degree $R$ used in the Chebyshev expansion of the Fermi-operator of the projected Hamiltonian ~\eqref{chebexp} is chosen to be 1250 since the width of the occupied spectrum is larger than that of a pseudopotential calculation. Just as in the case of the previous benchmark calculations, we use adaptive truncation tolerances ($\delta_c$, $\delta_l$) with  $\delta_l = 10^{4} \delta_c$  and vary $\delta_c$ as a function of relative change in the ground-state energy between two successive iterations ($\delta E_r$). We use the specific choice of truncation tolerance given in equation ~\eqref{delta_Al} with $C_1 = 10^{-11}$, $p = 2/3$ and $q=3/2$.  We note that much tighter tolerances have been used for $\delta_c$ in comparison to pseudopotential calculations in order to control the accumulation of truncation errors in applying a very high degree Chebyshev filter. In all our simulations, using the proposed method as well as ChFSI-FE, the SCF is terminated when $\delta E_r < 10^{-6}$ for three successive iterations.

\begin{table}[htbp]
\caption{\small{Comparison of ground-state energies for various sizes of silicon clusters.}}
\label{tab:energies}
 \begin{center}
 \begin{tabular}{|c|c|c|c|}
   \hline
Cluster & DoF & Proposed Method & ChFSI-FE   \\ \hline\hline
1x1x1 &$5,136,901$ & $-288.32016$ & $-288.32046$  \\ \hline
2x1x1 & $9,676,481$  &  $-288.33380$ & $-288.33411$ \\ \hline
2x2x1 & $16,358,791$ &  $-288.34715$ & $-288.34791$   \\ \hline
2x2x2 &$27,208,731$ &  $-288.35985$ & $-288.36140$\\ \hline
3x3x3 &$79,226,681$ &  $-288.37132$ & $-288.37341$\\\hline \hline
\end{tabular}
\end{center}\label{tab:siliconEnergies}
\end{table}
\vspace{-0.2in}
The ground-state energies computed using the proposed subspace projection algorithm for varying sizes of silicon nano-clusters considered in this work are tabulated in Table~\ref{tab:siliconEnergies}. These results indicate that the proposed method provides good accuracies in the ground-state energies, where the computed energies are within $3~mHa$ per atom of the reference energies computed using ChFSI-FE. The computational times for the full SCF convergence for varying sizes of silicon nano-clusters are plotted in figure~\ref{fig:siliconScaling}  for the proposed method, as well as, the reference calculations using ChFSI-FE. These results indicate that the proposed approach is computationally efficient, compared to ChFSI-FE, for system sizes beyond 18 atoms and provides a factor of $\sim 3$ speedup for the 3 $\times$ 3 $\times$ 3 nano-cluster containing  280 atoms. Using these computational times, the estimated scaling for the proposed approach is found to be approximately $\order (N^{1.85})$, while scaling in the case of ChFSI-FE is $\order (N^{2.21})$. The average computational time per SCF iteration is shown in figure~\ref{fig:siliconScalingPerSCF}, and the scaling is computed to be $\order (N^{1.75})$ for the proposed subspace projection method and $\order (N^{2.11})$ for ChFSI-FE. The deterioration in the scaling for all-electron calculations in comparison to pseudopotential calculations is due to tighter truncation tolerances employed in order to control the accumulation of the truncation errors during the application of a very high degree Chebyshev filter. Nevertheless, the proposed approach provides significant savings, which will increase with increasing system size. We refer to the Appendix for the scaling of various components of the subspace projection algorithm in the case of all-electron calculations and a discussion. Figure~\ref{fig:Sicontour} shows the electron density contours on the mid-plane of $3\times 3\times 3$ nano-cluster computed using the proposed subspace projection technique.


%% file: conclusions.tex
In the present study, we formulated a subspace projection technique in the framework of higher-order spectral finite-element discretization of the Kohn-Sham DFT problem in order to reduce the computational complexity involved in traditional solution approaches that compute the canonical orthonormal Kohn-Sham eigenfunctions. The proposed approach provides a single unified framework to handle both insulating and metallic materials system. Further, both pseudopotenial as well as all-electron calculations can be conducted using the proposed methodology.

The development of the proposed approach involved bringing together four main ideas. Firstly, we employed a higher-order spectral finite-element basis for the discretization of the Kohn-Sham DFT problem. The adaptive nature of the finite-element basis is crucial for efficiently handling all-electron DFT calculations. Secondly, we employed the Chebyshev filtering approach~\cite{saad2006} to directly compute an approximation of the occupied eigenspace in each SCF iteration. In this Chebyshev filtering step, we effectively exploited the finite-element structure to conduct matrix-vector products, associated with the action of the Chebyshev filter on a space of localized trial wavefunctions from previous SCF iteration, in linear-scaling complexity. We subsequently employed the localization procedure proposed by Garcia et al.~\cite{carlos} to compute atom-centered non-orthogonal localized basis (localized wavefunctions) spanning the Chebyshev filtered subspace. We employed an adaptive tolerance, where looser tolerances are used in initial iteration and progressively become tighter as the SCF approaches convergence, to provide a compact support for the localized wavefunctions. The use of adaptive tolerance provides strict control on the accuracy of the calculation, which is reflected in our benchmark calculations. Finally, we computed the projection of the Hamiltonian into the non-orthogonal localized basis, and used the Fermi-operator expansion~\cite{baer97} to compute the relevant quantities, including the finite-temperature density matrix, electron density and the band energy. We note that as the Fermi-operator expansion is computed in the projected subspace, the spectral width of the projected Hamiltonian is bounded---$\order(1\,Ha)$ for pseudopotential calculations and $\order(10\,Ha)$ for all-electron calculations---and can efficiently be computed for both pseudopotential and all-electron calculations. We demonstrated from complexity estimates that, for well-localized wavefunctions with a compact support, all operations in the proposed algorithm are linear-scaling in complexity.

The accuracy and performance of the proposed method was investigated on three different materials system: (i) a series of aluminum nano-clusters up to 3430 atoms representing a metallic system; (ii) a series of alkane chains up to 7052 atoms representing an insulating system; (iii) a series of silicon nano-clusters up to 3920 electrons representing a semiconducting system. Pseudopotential calculations were conducted on aluminum nano-clusters and alkane chains, whereas all-electron calculations were performed on silicon nano-clusters. In all the cases, the proposed method provided ground-state energies that are in excellent agreement with reference calculations, with accuracies commensurate with chemical accuracy. From these benchmark calculations, the computational complexity of the proposed approach was computed to be $\order (N^{1.46})$ for aluminum nano-clusters, $\order (N^{1.18})$ for the alkane chains, and $\order (N^{1.75})$ for the all-electron silicon nano-clusters. The deviation from linear-scaling, in practice, is due to the use of adaptive tolerances with tighter tolerances in the later SCF iterations in order to ensure strict control on the accuracy of the calculations. This affects the scaling due to reduced sparsity in the localized wavefunctions. We further note that using the proposed approach $\sim 10-$fold speedups were obtained with respect to reference benchmark calculations for the largest systems.

The present work demonstrates a methodology to conduct large scale electronic structure calculation using spectral finite-element discretizations at reduced scaling and presents an important direction for electronic structure calculations employing the finite-element basis. The computational efficiency as well as scaling in the case of all-electron calculations can be further improved by using a finite-element basis enriched with single-atom wavefunctions, or, alternately, the partitions-of-unity finite-element approach~\cite{PUFE}, and is currently being investigated. Also, the subspace projection method, which exploits the locality of the electronic structure, constitutes an important step in developing seamless coarse-graining techniques for Kohn-Sham density functional theory, in the similar spirit of quasi-continuum reduction techniques for electronic structure calculations~\cite{QCOFDFT,phanish2013a}.

%% file: acknowledgements.tex
We thank Janakiraman Balachandran for useful discussions on the implementation of non-local pseudopotentials in the finite-element framework. We gratefully acknowledge the support of Air Force Office of Scientific Research through Grant No. FA9550-13-1-0113 under the auspices of which the mathematical and computational framework was developed. We also gratefully acknowledge the U.S. Department of Energy, Office of Basic Energy Sciences, Division of Materials Science and Engineering under award number DE-SC0008637 that funds the Predictive Integrated Structural Materials Science (PRISMS) center at University of Michigan for providing the support to conduct the large-scale benchmark calculations presented in this work. V.~G. also acknowledges the Alexander von Humboldt Foundation through a research fellowship, and is grateful to the hospitality of the Max-Planck Institute for Mathematics in Sciences while completing this work. We also acknowledge Advanced Research Computing at University of Michigan for providing the computing resources through the Flux computing platform.

%% file: appendix.tex
\section{Scaling performance of individual components of the subspace projection technique}\label{scalingcomponent}
\subsection{Case study: Pseudopotential calculations}
We report scaling of individual components of the proposed subspace projection algorithm with system size for the benchmark calculations involving aluminum nano-clusters and alkane chains reported in section~\ref{sec:results}. The average CPU-times per SCF iteration of the various components involved in the proposed technique (cf. section~\ref{sec:fem})---namely: a) Chebyshev filtered subspace iteration (ChFSI) b) Localization procedure (Loc) c) Subspace projection in the non-orthogonal basis (SubProj) and d) Electron-density computation (ElecDen)---have been recorded and are plotted against number of atoms.
\setcounter{paragraph}{0}
\paragraph{Aluminum nano-clusters:} 

\vspace{0.2in}
\begin{figure}
\begin{center}
\includegraphics[width=0.5\textwidth]{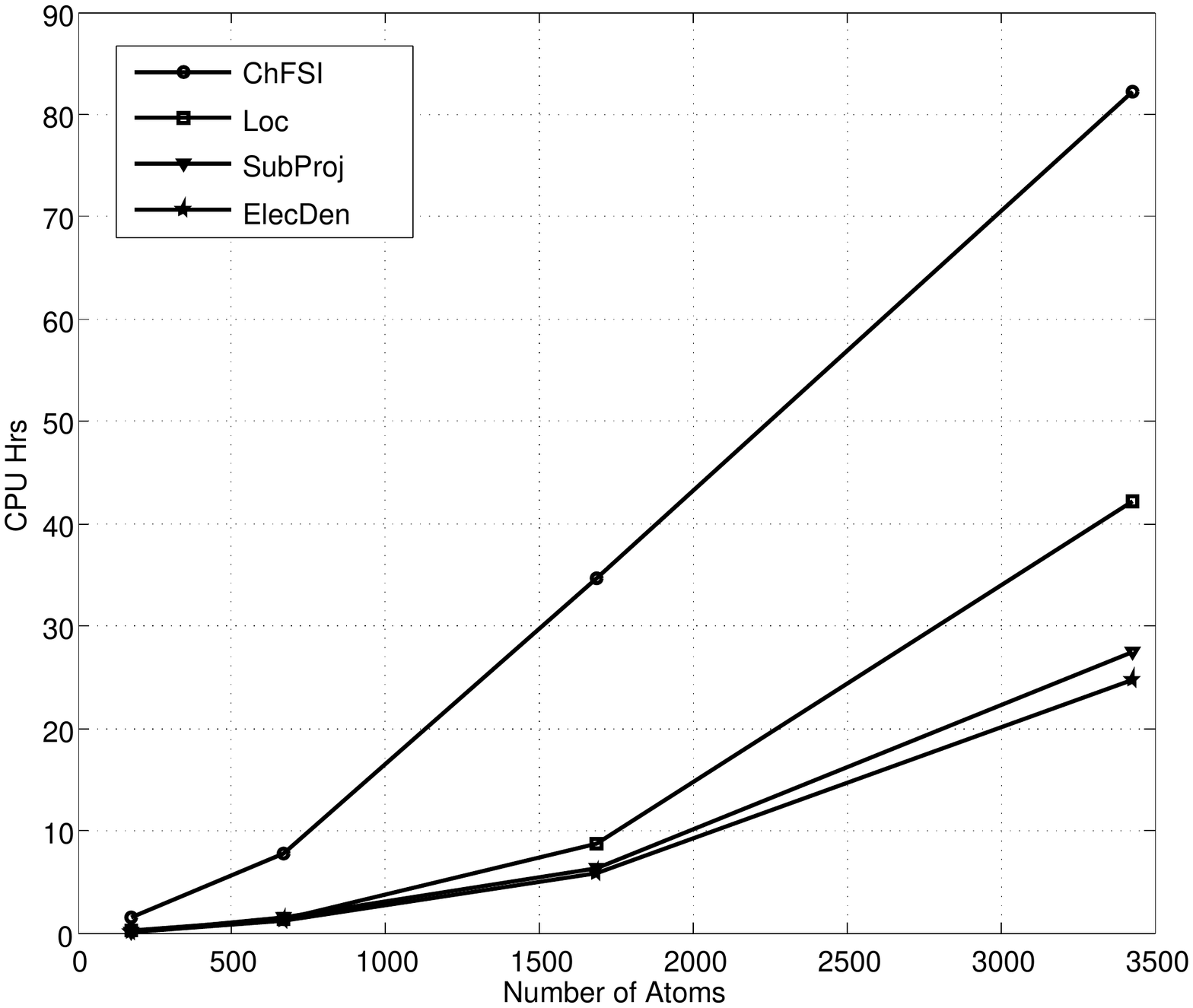}
\vspace{-1.0in}
\caption{\small{Average computational times per SCF iteration for individual components of the proposed method. Case Study: Aluminum nano-clusters.}}
\label{fig:alumComp}
\end{center}
\begin{center}
\includegraphics[width=0.5\textwidth]{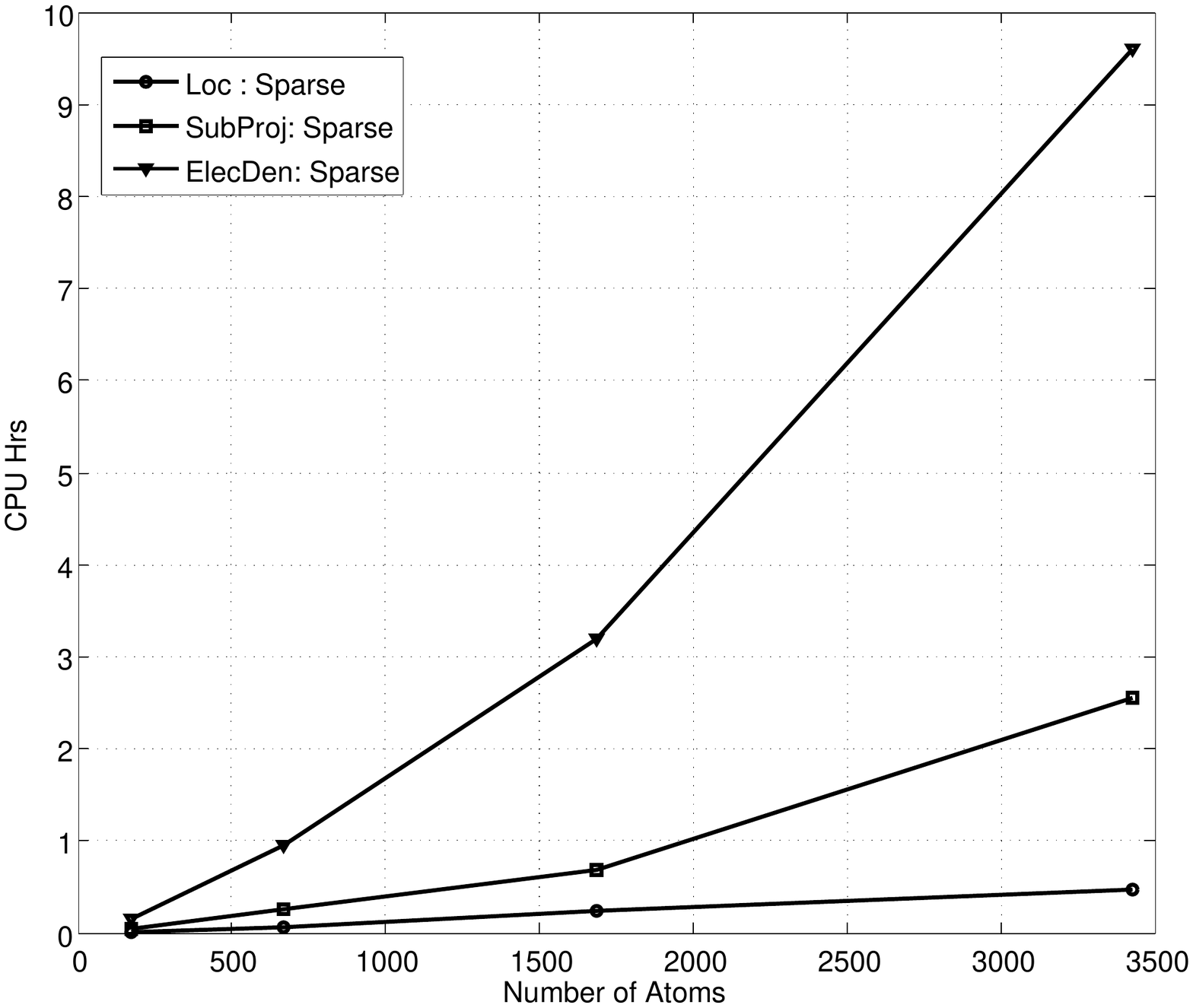}
\vspace{-1.0in}
\caption{\small{Average computational times per SCF iteration for individual components when sparse data-structures are active. Case Study: Aluminum nano-clusters.}}
\label{fig:alumCompSparse}
\end{center}
\end{figure}

\begin{figure}[htbp]
\begin{center}
\includegraphics[width=0.5\textwidth]{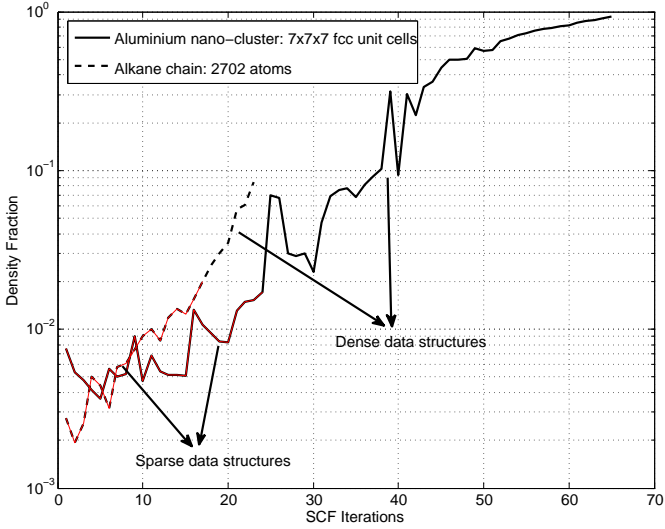}
\vspace{-1.0in}
\caption{\small{Variation of density fraction with SCF iteration showing where the sparse data-structures (brown line) and dense data-structures  are used. Case Study: Aluminum 7x7x7 nano-cluster and Alkane chain C$_{900}$H$_{1802}$.}}
\label{fig:denfrac}
\end{center}
\end{figure}

\begin{figure}[htbp]
\begin{center}
\includegraphics[width=0.5\textwidth]{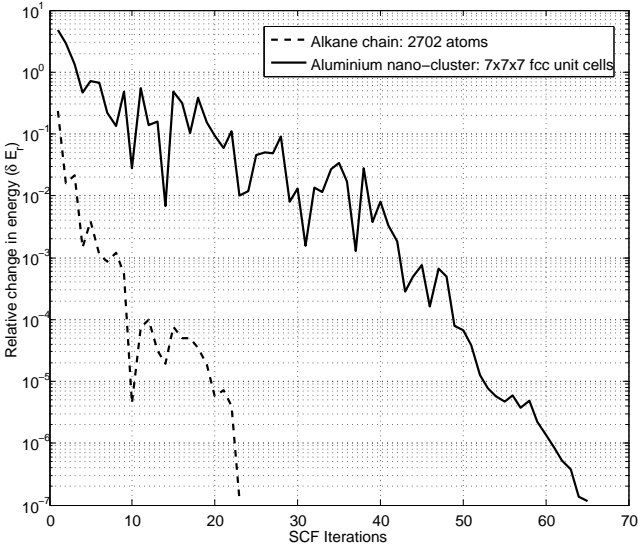}
\vspace{-1.0in}
\caption{\small{Relative change in ground-state energy with SCF iteration. Case Study:Aluminum 7x7x7 nano-cluster and Alkane chain C$_{900}$H$_{1802}$.}}
\label{fig:error}
\end{center}
\end{figure}

Figure ~\ref{fig:alumComp} shows the average computational CPU-times per SCF iteration for individual components of the subspace projection technique in the case of aluminum nano-clusters. These results indicate a computational complexity of $\order(N^{1.34})$ for the Chebyshev filtered subspace iteration, $\order(N^{1.91})$ for the localization procedure, $\order(N^{1.93})$ for the subspace projection and $\order(N^{1.94})$ for the computation of electron-density. The deviation of the scaling behaviour from linearity for the individual components of the algorithm is primarily due to the use of an adaptive tolerance (cf. equation~\eqref{delta_Al}) to truncate the localized wavefunctions, which is important to ensure the accuracy of the subspace projection technique. We recall that looser truncation tolerances are employed in the initial SCF iterations with progressive tightening of the tolerances during the course of SCF convergence. We note that the Chebyshev filtering procedure has the dominant computational cost, and also exhibits better scaling in comparison to the other components of the subspace projection algorithm. The better scaling behaviour of Chebyshev filtering procedure can be attributed to the matrix-vector multiplications performed at the finite-element level only if the relevant wave-functions have a non-zero value in the finite-element considered (cf. Section~\ref{Sec:ChebyshevFilter}). This allows us to naturally exploit the sparsity of the wavefunctions at the finite-element level even with tight tolerances. The higher scaling observed for the other components of the subspace projection algorithm is due to the use of dense data-structures of PETSc when the density fraction of the relevant matrices is above 2\%, as sparse PETSc data-structures have been observed to be efficient only when the density fraction is $< 1-2$\%. Figure~\ref{fig:alumCompSparse} shows the average computational CPU-times per SCF iteration when the sparse data-structures are active. The results indicate a computational complexity of $\order(N^{1.20})$ for the localization procedure, $\order(N^{1.32})$ for the subspace projection and $\order(N^{1.34})$ for the computation of electron-density. We remark that the scaling exponents of these individual components are closer to linearity when the sparse data-structures are active, but still deviate from linearity. We attribute this deviation from linearity to the delocalized nature of the wavefunctions for a metallic system. This delocalized nature of wavefunctions results in a higher density fraction (lesser sparsity) of truncated wavefunctions ($\bPhi_L$) for any given truncation tolerance in comparison to the insulating alkane chains (see discussion below on alkane chains). We also note that, as expected, the scaling of the individual components of the subspace projection technique (Loc, SubProj, ElecDen) is close to cubic-scaling ($\sim\order(N^{2.8})$) when dense data-structures are employed. Figure ~\ref{fig:denfrac} shows the variation of density fraction of $\bPhi_L$ with SCF iteration in the case of aluminum nano-cluster (7x7x7 cluster with 1688 atoms) highlighting the SCF iterations in which sparse data-structures are active. Figure~\ref{fig:error} shows the variation of relative error in ground-state energy with SCF iteration for the same benchmark problem.

\begin{figure}
\begin{center}
\includegraphics[width=0.5\textwidth]{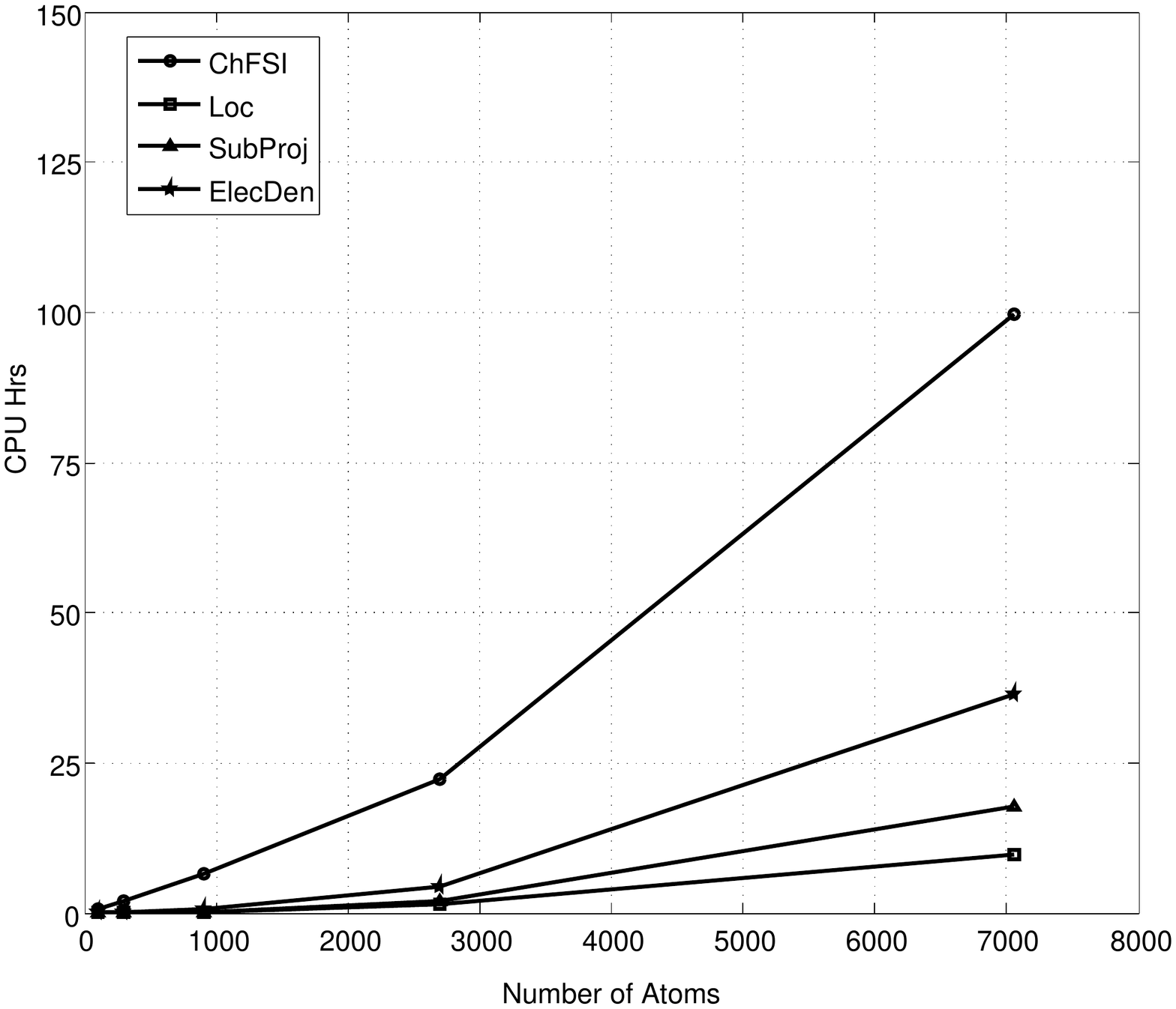}
\vspace{-1.0in}
\caption{\small{Average computational times per SCF iteration for individual components of the proposed method. Case Study: Alkane chain.}}
\label{fig:alkaneChain}
\end{center}
\begin{center}
\includegraphics[width=0.5\textwidth]{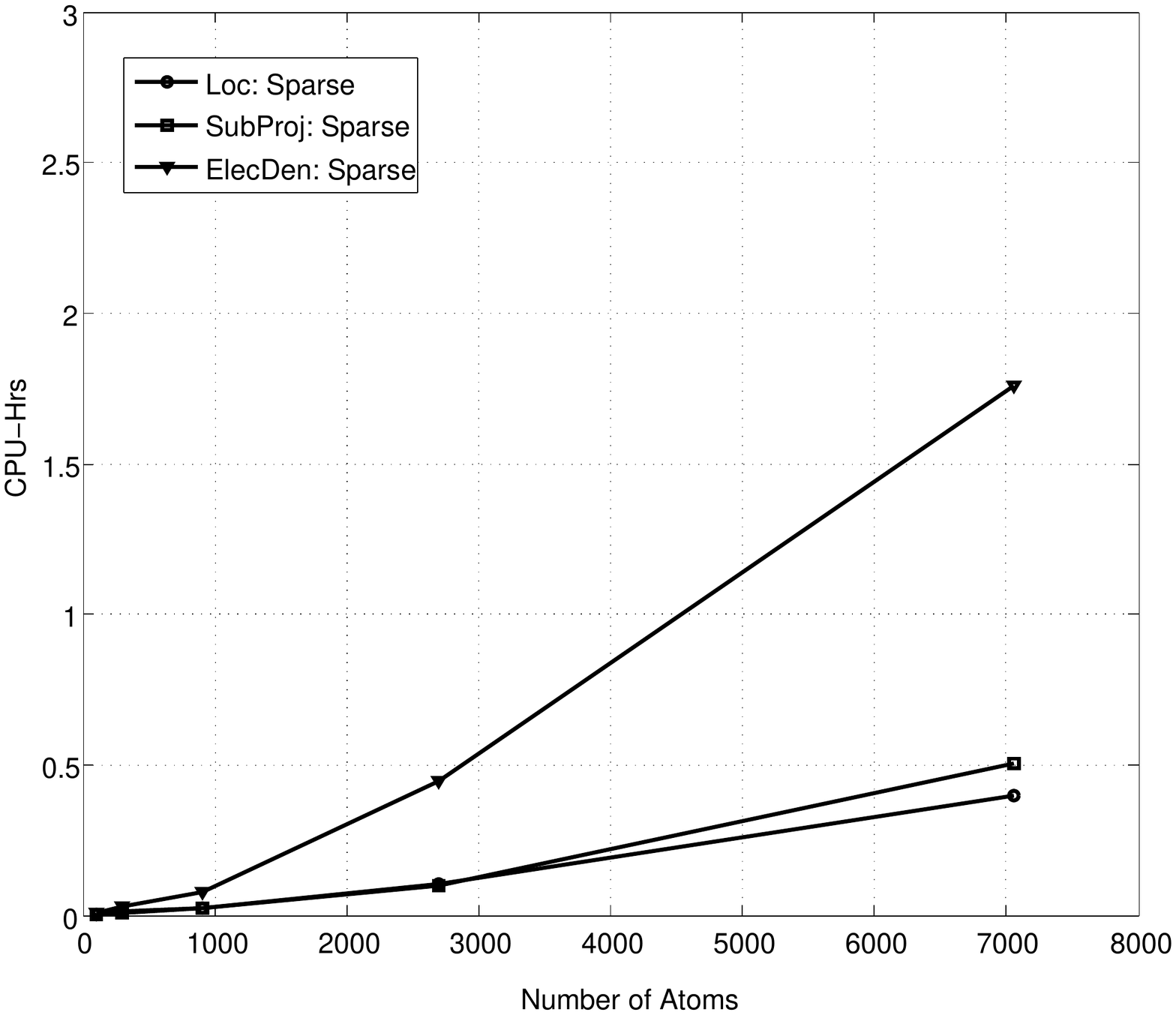}
\vspace{-1.0in}
\caption{\small{Average computational times per SCF iteration for individual components when sparse data-structures are active. Case Study: Alkane chain.}}
\label{fig:alkaneChainSparse}
\end{center}
\end{figure}

\paragraph{Alkane Chains:} 
Figure ~\ref{fig:alkaneChain} shows the average computational CPU-times per SCF iteration for individual components of the subspace projection technique in the case of alkane chains. These results indicate a computational complexity of $\order(N^{1.15})$ for the Chebyshev filtered subspace iteration, $\order(N^{1.80})$ for the localization procedure, $\order(N^{1.85})$ for the subspace projection and $\order(N^{1.91})$ for the computation of electron-density. As in the case of aluminum clusters, the Chebyshev filtering procedure comprises the dominant computational cost of the subspace projection algorithm, and is almost linear-scaling for this system. The improved scaling of the Chebyshev filtering procedure in comparison to aluminum nano-clusters can be attributed to better localization of wavefunctions in the case of alkane chains, an insulating material system. The higher scaling of other components of the subspace projection algorithm is once again due to the use of dense data-structures of PETSc when density fraction of the relevant matrices is above $2\%$, while sparse data-structures are employed only when the density fraction is $<2\%$. Figure ~\ref{fig:alkaneChainSparse} shows the average computational CPU-times per SCF iteration when sparse data-structures are active. The results indicate a computational complexity of $\order(N^{1.13})$ for the localization procedure, $\order(N^{1.25})$ for the subspace projection and $\order(N^{1.29})$ for the computation of electron-density. We remark that these scaling exponents are smaller in comparison to those of aluminum nano-cluster (metallic nature) due to the localized nature of the wavefunctions, thus resulting in smaller density fractions (greater sparsity) of the truncated wavefunctions ($\bPhi_L$). Figure~\ref{fig:denfrac} shows the variation of density fraction with SCF iteration in the case of C$_{900}$H$_{1802}$ (2702 atoms) highlighting the SCF iterations in which sparse data-structures are active. Figure~\ref{fig:error} shows the variation of relative error in ground-state energy with SCF iterations for the same problem.
\subsection{Case study: All-electron calculations}
\begin{figure}
\begin{center}
\includegraphics[width=0.5\textwidth]{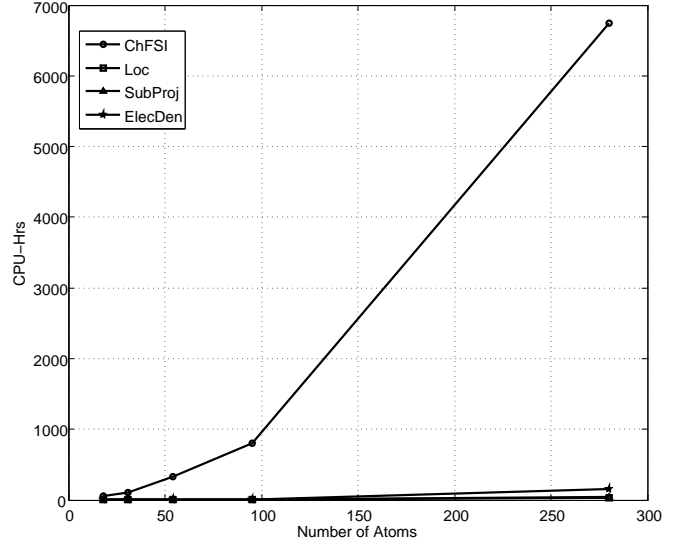}
\vspace{-1.0in}
\caption{\small{Average computational times per SCF iteration for individual components of the proposed method. Case Study: Silicon nano-clusters.}}
\label{fig:siCluster}
\end{center}
\end{figure}
\paragraph*{Silicon nano-clusters:} Figure ~\ref{fig:siCluster} shows the average computational CPU-times per SCF iteration for individual components of the subspace projection technique in the case of all-electron calculations performed on silicon nano-clusters. These results indicate a computational complexity of $\order(N^{1.74})$ for the Chebyshev filtered subspace iteration, $\order(N^{2.72})$ for the localization procedure, $\order(N^{2.55})$ for the subspace projection and $\order(N^{2.51})$ for the computation of electron density. In comparison to the pseudopotential calculations, the scaling of the individual components deviate significantly from linearity. The main reason for this significant deviation is due to the tighter adaptive tolerances employed in all-electron calculations. These tighter tolerances were necessary to avoid error accumulations during the Chebyshev filtering procedure as a very high order Chebyshev filter is needed in all-electron calculations to filter the large unwanted spectrum. For these tighter tolerances, the density fractions for the various relevant matrices were greater than $2\%$ even during the initial SCF iterations, which explains the observed close to cubic-scaling of the localization procedure, subspace projection, and computation of electron density. However, we note that the significantly dominant cost for all-electron calculations is the Chebyshev filtering step, which naturally exploits the sparsity in truncated wavefunctions, even for high density fractions, by computing the matrix-vector products at the finite-element level only if the relevant wave-functions have a non-zero value in the finite-element being considered (cf. Section~\ref{Sec:ChebyshevFilter}). Thus, the overall scaling of the proposed technique for all-electron calculations is determined by the scaling of the Chebyshev filtering step, even for modestly sized systems. While, the performance of the proposed approach for all-electron calculations is not as good as the performance for pseudopotential calculations, the proposed approach does offer significant computational savings ($\sim 3$ fold in comparison to ChFSI-FE for the silicon nano-cluster containing $\sim 4000$ electrons). Further, we note that the need of a very high degree Chebyshev filter for all-electron calculations can be mitigated by employing an enriched finite-element basis, where the finite-element basis functions are enriched by numerically computed single-atom wavefunctions, and this can potentially lead to better overall scaling of the method for all-electron calculations.